\documentclass[%
 reprint,
 amsmath,amssymb,
 aps,
pre,
]{revtex4-1}
\usepackage{graphicx}
\usepackage{dcolumn}
\usepackage{amscd}
\usepackage{amsmath}
\usepackage{array}
\usepackage{color}
\usepackage{amsbsy}
\usepackage{mathrsfs}
\usepackage{mathtools}
\usepackage{bm}
\usepackage{changepage}

\begin{document}



\title{Funneled Potential and Flux Landscapes Dictate the Stabilities of both the States and the Flow: Fission Yeast Cell Cycle} 

\author{Xiaosheng Luo}
\affiliation{Applied Science Department at Little Rock, University of Arkansas,\\ Little Rock, AR 72204}
\author{Liufang Xu}
\affiliation{Department of Physics and Biophysics $\&$ Complex System Center, Jilin University Changchun 130012, P. R. China}
\author{Jin Wang}%
 \email{jin.wang.1@stonybrook.edu (Corresponding Author)}
\affiliation{Department of Physics and Biophysics $\&$ Complex System Center, Jilin University Changchun 130012, P. R. China}
\affiliation{
 State Key Laboratory of Electroanalytical Chemistry, Changchun Institute of Applied Chemistry Chinese Academy of Sciences, Changchun, Jilin 130022, China
}%
\affiliation{
Department of Chemistry and Physics
State University of New York at Stony Brook,\\
Stony Brook, NY 11794 \\
}%


\begin{abstract}

Using fission yeast cell cycle as an example, we uncovered that the non-equilibrium network dynamics and global properties are determined by two essential features: the potential landscape and the flux landscape. These two landscapes can be quantified through the decomposition of the dynamics into the detailed balance preserving part and detailed balance breaking non-equilibrium part. While the funneled potential landscape is often crucial for the stability of the single attractor networks, we have uncovered that the funneled flux landscape is crucial for the emergence and maintenance of the stable limit cycle oscillation flow. This provides a new interpretation of the origin for the limit cycle oscillations: There are many cycles and loops existed flowing through the state space and forming the flux landscapes, each cycle with a probability flux going through the loop. The limit cycle emerges when a loop stands out and carries significantly more probability flux than other loops. We explore how robustness ratio (RR) as the gap or steepness versus averaged variations or roughness of the landscape, quantifying the degrees of the funneling of the underlying potential and flux landscapes. We state that these two landscapes complement each other with one crucial for stabilities of states on the cycle and the other crucial for the stability of the flow along the cycle. The flux is directly related to the speed of the cell cycle. This allows us to identify the key factors and structure elements of the networks in determining the stability, speed and robustness of the fission yeast cell cycle oscillations. We see that the non-equilibriumness characterized by the degree of detailed balance breaking from the energy pump quantified by the flux is the cause of  the energy dissipation for initiating and sustaining the replications essential for the origin and evolution of life.  Regulating the cell cycle speed is crucial for designing the prevention and curing strategy of cancer.

\end{abstract}


\maketitle 


\section*{Author Summary}
We have uncovered that the non-equilibrium network dynamics and global properties are determined by two essential features: the potential landscape and the flux landscape. We have found that the funneled potential landscape is crucial for the stability of the states on the cell cycle, however, the stabilities of the oscillation states cannot guarantee the stable directional flows. We have uncovered that the funneled flux landscape is important for the emergence and maintenance of the stable limit cycle oscillation flow. This work will allow us to identify the key factors and structure elements of the networks in determining the stability, speed and robustness of the fission yeast cell cycle oscillations. We see that the non-equilibriumness characterized by the degree of detailed balance breaking from the energy pump quantified by the flux is the cause of  the energy dissipation for initiating and sustaining the replications essential for the origin and evolution of life. Regulating the cell cycle speed is crucial for designing the prevention and curing strategy of cancer.

\section*{Introduction}
The global stability and robustness are crucial for maintaining the function. They are also important for uncovering underlying mechanisms of the networks.\cite{Jackson, Frauenfelder, Wolynes, Wang2003PRL, Haken, Graham, Sasai} However, it is difficult to quantify them for dynamic systems and networks. This presents a challenge for the dynamical systems and the field of systems biology.\cite{Qian2006, Qian2009, Wang2007PLOS, Wang2007BJ,Wang2008PRE, Wang2008PNAS,Wang2008PNAS2, WangAdvancePhysics2015, WangChinesePhysicsBReview2016, WangZhangWangJCP2010, FengWangJCP2011,WangZhangXuWangPNAS2011, WangLiWangPNAS2010, ZhangXuZhangWangWangJCP2012, LiWangPNAS2014, LiWangJRSI2014}

In equilibrium systems, the global nature of the system is
characterized by the underlying equilibrium potential landscape $U$ which is directly linked to the equilibrium probability through the Boltzmann distribution law $P\sim exp(-\beta U)$. The local dynamics is determined by the gradient of the equilibrium potential landscape. However, most dynamical systems do not typically have a gradient potential as in the equilibrium case. They are open systems usually not in isolations. Global natures of such systems are hard to address. In addition, for mesoscopic systems, the intrinsic fluctuations can also be significant. Under stochastic fluctuations, instead of following the dynamical trajectories which are stochastic and unpredictable, the evolution of the probabilistic distributions should be followed, which is inherently global as well as predictable due to its intrinsic linearity. The probabilistic evolution is governed by the master equations for discrete state space (more general) and Fokker-Planck equations for continuous state space.

It turns out the steady state distribution of the probability evolution in long time limit can give a global quantification of the dynamical systems\cite{Haken, Graham, Sasai, Qian2006, Qian2009, Wang2007PLOS, Wang2007BJ,Wang2008PRE, Wang2008PNAS,Wang2008PNAS2, WangAdvancePhysics2015, WangChinesePhysicsBReview2016, WangZhangWangJCP2010, FengWangJCP2011,WangZhangXuWangPNAS2011, WangLiWangPNAS2010, ZhangXuZhangWangWangJCP2012, LiWangPNAS2014, LiWangJRSI2014}. This defines a probability or weight landscape for characterizing the system states. On the other hand, the dynamics of the systems can be decomposed to gradient of the potential landscape related to the steady state probability distribution and a curl probability flux. The existence of a non-zero curl flux directly reflects the degree of the breakdown of the detailed balance. This quantifies the degree of the non-equilibrium. While this decomposition is shown explicitly in continuous space through Fokker-Planck equation description of the stochastic dynamics \cite{Wang2007PLOS, Wang2007BJ,Wang2008PRE, Wang2008PNAS,Wang2008PNAS2, WangAdvancePhysics2015, WangChinesePhysicsBReview2016, WangZhangWangJCP2010, FengWangJCP2011,WangZhangXuWangPNAS2011, WangLiWangPNAS2010, ZhangXuZhangWangWangJCP2012, LiWangPNAS2014, LiWangJRSI2014}, the corresponding decomposition and associated statistics of stochastic dynamics in discrete space from the master equation still needs further explorations \cite{Wang2007BJ,Wang2008PRE, ZhangWangJCP2014, ZhangWangNewJPhys2015,Schnakenberg, Qian, Ouya06,Tang06,Zia, Ge}.

In this work, we study the more general stochastic dynamics in discrete space of the non-equilibrium networks (Markov chains) governed by the probabilistic master equation. We found the network dynamics and global properties are determined by two features: the potential landscape and the probability flux landscape. While potential landscape quantifies the probabilities of different states forming hills and valleys, the probability flux landscape is composed of many flux loops flowing in state space. Therefore, statistics of the flux loops becomes important. These two landscapes can be quantitatively constructed through the decomposition of the dynamics into the detailed balance part and non-detailed balance part.

We found that while funneled landscape is crucial for the stability of the single
attractor networks and stability of oscillation states.  The funneled flux landscape is crucial for maintaining the stable limit cycle oscillation flow. The stability and the robustness of the networks can be quantified through a dimensionless ratio of the gap or steepness versus the averaged variations or roughness of the landscape (which measures the degree of funnel, we termed as robustness ratio RR), and explored under the changes of the network topologies and stochastic fluctuations.

This flux landscape picture  provides a new interpretation of
the origin of the limit cycle oscillations. The global oscillation only
emerges when one specific loop stands out and carries much more
probability flux, and therefore becomes more probable than the rest of the
others.

We specifically studied the fission yeast cell cycle as an example to
illustrate the idea. We found the flux landscape of the fission
yeast cell cycle oscillations is funneled, which guarantees its stability
and the robustness of the oscillation flows. The global stability is quantified by the
robustness ratio RR of the funneled flux landscape and the robustness is quantified by RR against the changes in topology of the network (wirings) and stochastic
fluctuations. The flux is directly related to the speed of the cell cycle. The landscape analysis here allows us to identify the key factors and structure elements of the networks in determining the global stability, speed, and robustness of the fission
yeast cell cycle oscillations.  We see that the non-equilibriumness characterized by the degree of detailed balance breaking from the energy pump quantified by the flux is the cause of  the energy dissipation for initiating and sustaining the replications essential for the origin and evolution of life. The cell cycle speed is a hallmark of cancer. Regulating the cell cycle speed thus provides a possible strategy for preventing and curing strategy against cancer.

\section*{Model and Methodology}
\subsection*{The stochastic boolean model of fission yeast cell cycle}
\begin{figure*}[htp]
\includegraphics[width=1.0\textwidth]{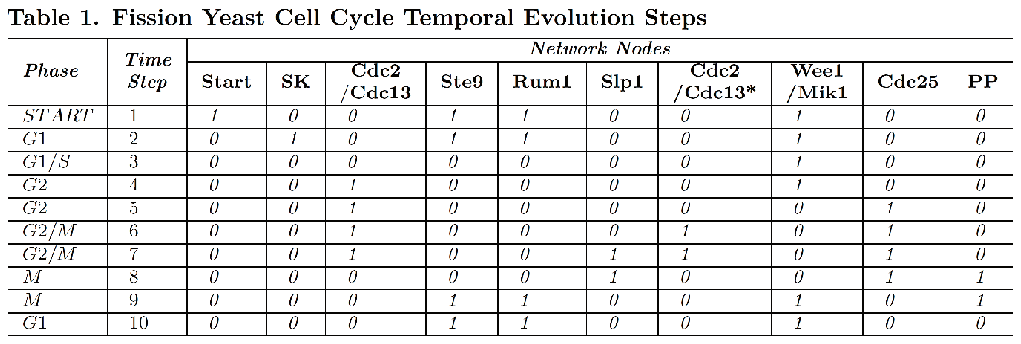}
\label{table1}
\end{figure*}

\begin{figure}[htp]
\includegraphics[width=0.45\textwidth]{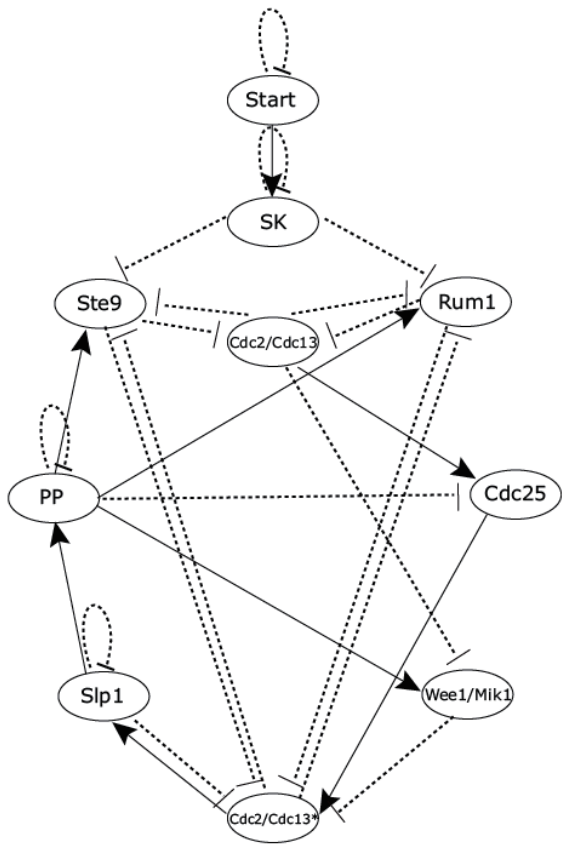}
\caption{The cell cycle regulatory network of fission yeast with 10 gene nodes. The solid arrow ($\rightarrow$) represents positive regulations between gene nodes. The inhibition sign ($\cdot\cdot\cdot|$ )represents negative regulations between gene nodes. }\label{fig:wiring}
\end{figure}

We follow a boolean network model mainly built for fission yeast in \cite{davidich2008boolean}. As illustrated in the Table \ref{table1}, the cell cycle period of fission yeast is divided into several phases: G1 phase $\rightarrow$ S phase $\rightarrow$ G2 phase $\rightarrow$ M phase $\rightarrow$ G1 phase. The network wiring diagram with 10 gene nodes is drawn in Fig.~\ref{fig:wiring}, therefore in the global state space, there are $2^{10}$ states. Each state is the combination of the "on" ($s_i=1$) and "off" ($s_i=0$) states of the 10 gene nodes of \textbf{Start, SK, Cdc2/Cdc13, Ste9, Rum1, Slp1, Cdc2/Cdc13*, Wee1/Mik1, Cdc25}, and \textbf{PP}, represented by a state vector  $ S=\{s_1, s_2, s_3,...,s_{10}\}$. With this representation, it can form a state space of a complex gene regulation interaction networks\cite{Kauffman}.  In general, one can use the boolean dynamics model to explore the coarse grained dynamics with the information of the wiring topology of the networks\cite{Tang04}.

As known, there are often intrinsic and extrinsic fluctuations in the cellular environments. We have to find an approach to probabilistically describe the network dynamics. Based on the Markovian process theory\cite{WangBJ2007, WangPRE2008, Free83, Davi93, Ouya06}, we firstly inherit the usage for the discrete state of on ($s_i=1$) and off ($s_i=0$) state for each gene expression level in the fission yeast network. Then we need to figure out the transition probability between two neighbor states in the Markov time series chain, which can be written as the product of each node transition\cite{Tang06,Wang2007BJ}
\begin{equation}\label{eq:1}
 T\Big( S(t')=\{s_1(t'),s_2(t'),...,s_{10}(t')\} \Big| S(t)\Big)
 =\prod^{10}_{i=1}T\Big(s_i(t') \Big| S(t)\Big),
\end{equation}
where $t$ and $t'$ are two close neighborhood time moments. Considering the simplifications of the interactions between two nodes (activated (+1) or repressed (-1)),  one can obtain the probabilities of transition or switching between states, given as\cite{Hopf82,Wang2007BJ,Ge}:
\begin{equation} \label{eq:2}
   T\big(s_i(t')|S(t)\big)=\left\{
   \begin{array}{ll}
   \frac{1}{2}+\frac{1}{2}tanh[\mu I],
   \quad &\textrm{if $I\neq 0$ and $s_i(t')=1$,} \\
   \frac{1}{2}-\frac{1}{2}tanh[\mu I],
   \quad &\textrm{if $I\neq 0$ and $s_i(t')=0$,} \\
   \gamma, \quad &\textrm{if $S(t)=G1$ and $i=0$,}\\
   1-c,
   \quad &\textrm{otherwise,}
   \end{array}\right.
\end{equation}
where $I=\sum^{11}_{j=1}a_{ij}S_j(t)$ is defined as the total input to a gene node from other nodes $s_j$ through connection strengths $a_{ij}$. Transition rate is determined by the input. Positive overall input tends to activate the state ($S_i(t')=1$), while negative overall input tends to suppress the state ($S_i(t')=0$). $\mu$ can be considered as transition rate or strength from the input to output of a gene or protein node, which is also related to the inverse of the fluctuation or noise strength. $\gamma$ represents the strength of the positive feedback or stimulation from the checkpoint to the \textbf{Start} when $G1$ is reached, and $c$ is a  parameter to quantify the effect of self-degradation  when the total input to a node is zero.

In this way, we obtain the evolution equation to guide the probabilistic dynamics, which is so called master equation\cite{Wang2007BJ, VanKampen, Gardiner}:
\begin{equation}\label{eq:3}
\frac{dP_i}{dt} = -\sum_jT_{ij}P_i + \sum_jT_{ji}P_j,
\end{equation}
where $P_i$ represents the probability of state $i$, and $T_{ij}$ represents the transition probability from state $i$ to state $j$. The physical meaning of the master equation is the conservation law of probability: the local change of the probability of a particular
state $i$ in time is equal to the probability flow (flux) from the other states to this state $i$ given by $\sum_jT_{ji}P_j$ subtracting the probability flow (flux) from the state $i$ to other states $\sum_jT_{ij}P_i$. By solving the $2^{10}=1024$ master equations numerically, we obtain both the time-dependent evolution and the steady-state probability of each state in the global state space.

\subsection*{Decompositions of boolean dynamics and probability flux loops}

For steady state, we set the left term of the master equation (\ref{eq:3}) to zero, that is $\frac{dP_i}{dt}=0$, then we obtain the numerical steady state solution $P^{(ss)}_{i}$, which is the long time limit. Given the steady solution, we can define the steady state flux between state i and j as: ${F_{ij}}^{ss}=-T_{ij} P^{(ss)}_{i} + T_{ji}P^{(ss)}_{j}$, If for any $i,j$ pair, ${F_{ij}}^{ss}=0$, this Markov chain is detail-balance preserved, and the steady state of the system becomes the equilibrium state (without net local flux), since $dP_i/dt=\sum_j{F_{ij}}^{ss}=0$. However, in general the steady state probability can be obtained, but it does not have to satisfy the detailed balance condition(${F_{ij}}^{ss}\neq0$). In other words, the net flux does not have to be zero. The system is then in
non-equilibrium steady state. Although the steady-state distribution is fixed and does not change in time, there can be an internal probability flow among states.

In order to study the non-equilibrium steady states and
characterize the global properties, one can separate the dynamical
process into two parts, a detailed balance part and a pure
irreversible non-detailed balance flux part by decomposing the
transition probability matrix $M$ \cite{Qian}. The master equation
can be rewritten as $dP/dt=M^TP$, where $P$ is the vector of
probability of all the discrete states, $M$ is the transition
probability matrix (or rate matrix) with $M_{ij}=T_{ij}, i\neq j$
and $M_{ii}=(-1)\sum_jT_{ij}$. We define a matrix $C$ such that
the $i$th row and $j$th column of it is given as
$C_{ij}=max\{T_{ij}P^{(ss)}_i-T_{ji}P^{(ss)}_j, 0\}/P^{(ss)}_i,
i\neq j$ and $C_{ii}=(-1)\sum_jC_{ij}$, and matrix $D$ whose $i$th
row and $j$th column is given as $D_{ij}=min\{T_{ij}P^{(ss)}_i,
T_{ji}P^{(ss)}_j\}/P^{(ss)}_i, i\neq j$ and
$D_{ii}=(-1)\sum_jD_{ij}$. It follows that $M=C+D$ and
$D^TP^{(ss)}=0$\label{eq:4}. Since $M^TP=(C+D)^TP^{(ss)}=0$,
$C^TP^{(ss)}=0$\label{eq:5}. By separating the transition
probability matrix this way, two Markov processes are obtained
\cite{Qian}. The probability transition matrix (or rate matrix)
M for characterizing the dynamics can be decomposed into two terms: C and D. Both C and D have the same steady-state(stationary) probability distribution, and one of the processes D satisfies detailed balance($D_{ij}{P_i}^{ss}=D_{ji}{P_j}^{ss}$), while the other C is non-detailed balanced and irreversible (if $C_{ij}{P_i}^{ss}>0$,$C_{ji}{P_j}^{ss}=0$). In this way, the dynamics is decomposed to detailed balance preserving part and detailed balanced breaking part.

The non-equilibrium irreversible part can be termed as the
circulation or flux part, since it can be further decomposed into
flux circles or loops with a flux value on each cycle
\cite{Qian}. The prove of the circulation also provides a way to
obtain all the circles and their corresponding flux values for the
dynamic part. By definition of flux, we have
$F_{ij}=-C_{ij}P^{(ss)}_{i} + C_{ji} P^{(ss)}_{j}=-F_{ji}$. Now
define $J_{ij}=C_{ij}P^{(ss)}_{i}, i\neq j; J_{ii}=0$, we have
$\sum_iJ_{ij}=\sum_iJ_{ji}$\label{eq:6}. Since $J_{ii}=0$, suppose
$J_{k_0k_1}>0$, from the summation equation just mentioned, we can
find a $k_2 \neq k_0, k_1$ such that $J_{k_1k_2}>0$. We can keep
on doing this, until a repeat is found: $k_n\in \{k_0, k_1,...,
k_{n-2}\}$. Suppose $k_n=k_{n_0}$, let $i_1=k_{n_0},
i_2=k_{n_0+1},...,i_{n_1}=k_{n-1}$ and $i_{n_1 +1}=i_1$, we now
construct a cycle or closed loop with $i_1,...,i_{n_1}$. Let
$r_1=min_{k=1,2,...,n_1}\{J_{{i_k}i_{k+1}} \}$, define $r_1$ as
the flux value of this cycle. Then subtract their flux value from
the whole J matrix, $J$, thus $J^{(1)}_{ij}=\left \{
\begin{array}{ll}
J_{ij}-r_1, i\in \{i_1,...,i_{n_1}\},\\
J_{ij}, otherwise.
\end{array}\right.$

If $J^{(1)}\neq 0$, repeat what we
did above to find another cycle as well as its flux value, then
subtract those fluxes from $J^{(1)}$ to get $J^{(2)}$. Since the
number of non-zero elements in $J^{(i)}$ is at least one less than
that in $J^{(i-1)}$, there exits an integer $N$ such that
$J^{(N+1)}=0$ (all the elements of the J matrix are zero).
Therefore for the non-detailed balanced part of the dynamics, the flux
can be decomposed to finite number of circles or closed loops,
each with a flux value, $J=\sum_{i=1}^{i=N} J^{(i)}$ \cite{Qian}.
Therefore, the decomposition and associated flux statistics can be directly carried out at the master equation level.

 On the other hand, one can also find another way for decomposition and associated statistics of the fluxes from the stochastic boolean trajectories. One can follow the trajectory from one state $S(t)$ at moment $t$ to another state $S(t')$ at the next moment $t'$. If one finds the same state at different moment, that is $S(t'')=S(t)$, then all the states between these two same states can be defined as one loop. Then one should remove this loop from the trajectory and repeat the steps above to obtain all the loops. By making statistical analysis on all the flux loops, one can calculate the flux landscape using the formula below, $U_{flux}=-ln{P_{flux}}$, where $P_{flux}=Flux_{loop}=lim_{N\rightarrow\infty}\frac{N_{J^{(i)}}}{N_{J}}$. \cite{Qian, Tang06, Ge, ZhangWangJCP2014, ZhangWangNewJPhysics2015}

Therefore we have two quantitative features to characterize the system, one is the steady state probability and the other is the non-zero steady state probability flux which can be further decomposed into loops. The steady state probability obeys the evolution equation of the transition probability matrix (or rate matrix) at long time limit characterizing only the detailed balance part. The detailed balance condition allows one to identify the path independent probability measures \cite{Zia}. This naturally leads to the potential. We can see how both potential and flux landscape influence the dynamics and stability of the system through an example on fission yeast cell cycle.

\subsection*{Entropy production rate and dissipation in the fission yeast boolean network}

As known, when an open system is under long time evolution, it can reach non-equilibrium steady state (NESS) \cite{Haken, Schnakenberg, Graham, Wang2007BJ,Wang2008PRE, Wang2008PNAS, WangAdvancePhysics2015, WangChinesePhysicsBReview2016, Qian07}. The local steady state flux ${F_{ij}}^{ss}=-T_{ij} P^{(ss)}_{i} + T_{ji}P^{(ss)}_{j}$ is not necessarily equal to zero (no detailed balance). In this condition we can define a generalized force referring to the generalized chemical potential (from $j$ to $i$) $A_{ji} = ln ( \frac{T_{ji} P_j}{T_{ij} P_i} $ ) \cite{Qian2006, Schnakenberg, Qian07, Wang2007PLOS, Wang2007BJ,Wang2008PRE, Wang2008PNAS,WangAdvancePhysics2015, WangChinesePhysicsBReview2016}. There is a mapping between the cellular networks and electric circuits. The flux $F_{ij}$ corresponds to current \textbf{I} and chemical potential $A_{ij}$ corresponds to voltage \textbf{V}. The non-equilibrium
cell network dissipates energy just as the electric circuits.

In the steady state, the heat loss rate is related to the entropy production rate. The entropy production or dissipation characterizes ``time irreversibility'' and provides a lower bound
for the actual heat loss in Boolean network\cite{Qian2006, Qian07, Wang2007PLOS, Wang2007BJ, Wang2008PRE, Wang2008PNAS,  WangAdvancePhysics2015, WangChinesePhysicsBReview2016}. The total
entropy change is equal to the part from the system or source plus the part from the bath or sink (dissipation). Since in steady state the entropy change of the system is equal to zero, thus the total entropy change (source) is equal to the entropy change of
the sink (dissipation). The total entropy change (source)  $= \sum
F_{ij} A_{ij} $ is the entropy production and the sink term is
dissipation. Therefore in steady state,  knowing the entropy
production, we know the dissipation quantitatively. The entropy
$S$ from the system part is defined as $S=-\sum_i P_i ln P_i $ and
entropy production rate $ \frac{dS_{tot}}{dt}$ is given by:
$\frac{dS_{tot}}{dt}= \sum F_{ji} A_{ji} = \sum_{ij} T_{ji} P_j ln
(\frac{T_{ji} P_j}{T_{ij} P_i}) $.

Entropy production is correlated with flux. When the steady state flux is zero, the entropy production or dissipation at steady state is zero. When the flux increases, the entropy production typically increases. Therefore, the entropy production or dissipation can also serve as a quantitative measure of how far away the system is from the equilibrium, or in other words, the degree of the detailed balance breaking.

\section*{Results and Discussions}

\subsection*{Potential Landscape of Fission Yeast Cell Cycle}
\subsubsection*{Quantifying the potential landscape}
\begin{figure}[htp]
\includegraphics[width=0.45\textwidth]{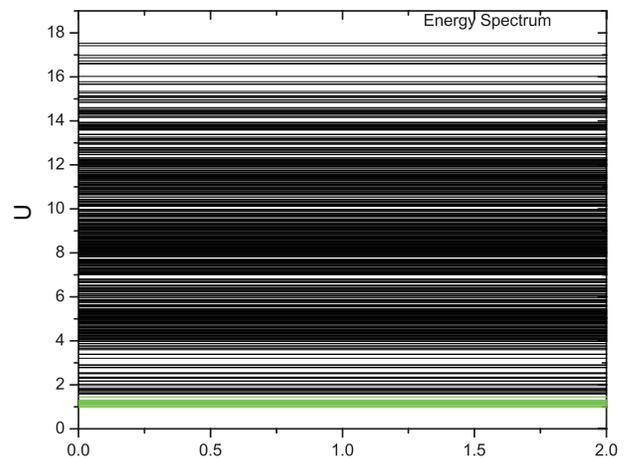}
\caption{Potential landscape spectrum  U of the $2^{10} $ states, where $\mu = 5$, $c = 0.001$ and
 $\gamma= 60\%$. The potential values of those 10 states of the biological pathway are in green lines. They are lower than the rest of the states.}\label{fig:espect}
\end{figure}

 To explore the global quantification of dynamical systems of this fission yeast networks, we define the underlying potential landscape $U$ from the steady state probability $U=-ln(P_{ss})$. To visualize the potential landscape $U$ over the $2^{10}$ states space, we firstly draw the landscape spectrum, which is shown in Fig.~\ref{fig:espect}. Then we define the Robustness Ratio (RR) as the potential gap between the lowest potential of the states and the average potentials of the rest of the states, versus the average variations measured by the standard deviation of potentials. That is $RR=\delta U/\Delta U$ where $\delta U= |U_m-<U>|$ and $\Delta U=\sqrt{<U^2>-<U>^2}$. The gap represents the separation between the lowest potential state with the rest of the decoys. This definition of RR holds for a single attractor. When we deal with cycle oscillations as fission yeast, we are interested in the stability of not only one state but the whole oscillating cycle states. Therefore we extend our definition to include all the states ($10$ states in our example of fission yeast cell cycle) on the oscillation path as the "native" states. So, for the cell cycle, the gap quantifies the bias or slope of the underlying landscape towards the native (potentials of all the states on the oscillation paths basin of attraction) while the average variations measure the fluctuations or roughness of the underlying landscape. In this way, we can quantify the topography of underlying potential landscape.

 Fig.~\ref{fig:espect} was shown with the parameters $\mu=5$, $c = 0.001$ and $\gamma=60\%$ for the fission yeast cell cycle model(Equ.~(\ref{eq:2})). In the potential landscape spectrum there are  $10$ states representing the biological cell cycle  phases (G1, S, G2, M) which are marked as green lines. We notice that they all settle at the bottom. This means the $10$ states  are sitting on the most stable cell cycle pathway. We then calculate $RR=2.29813$. $RR$ is significantly larger than $1$, which indicates the $10$ states along the biological cell cycle is separated from the others. So we state that this can be the reason that the cell cycle states are stable since the states of the biological path are all at low potentials and high probabilities sufficiently separated from others.

\begin{figure}[htp]
\includegraphics[width=0.45\textwidth]{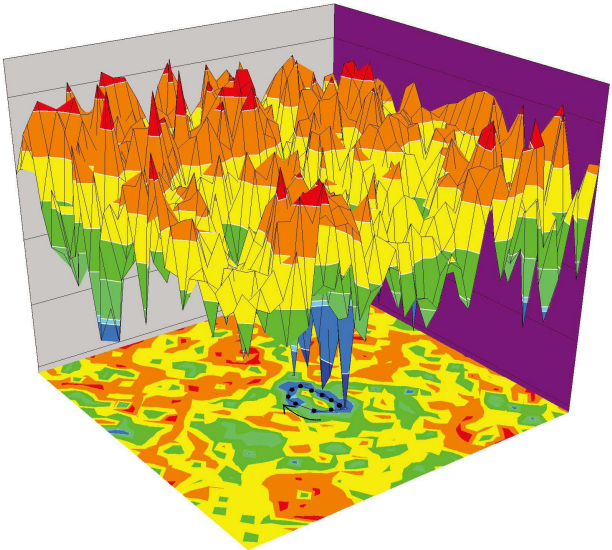}
\caption{Three dimensional potential landscape and two dimensional
contour in projected 2 dimensional state space. The vertical axis
and color represent the potential level of each state in the
three dimension and the contour map laying on the bottom respectively. The low
potential  valley of the potential is a cycle or closed ring,
which is exactly the biological cycle path with low potential
level, and this can also be seen more clearly on the contour map.}\label{fig:landscape}
\end{figure}

Furthermore, we draw all  those   states ($2^{10}=1024$) on a 2-dimensional surface by minimizing the distances between two states which have the strongest connections, and using the underlying potential energy to be the vertical axis. The color represents the potential level of each state on both surface and the bottom contour map. We obtain Fig.~\ref{fig:landscape}, in which we notice that the potential landscape shows a distinct topology with a Mexican hat like shape. This gives us a visualized picture for the potential landscape in $3D$. The valleys of this landscape correspond to exactly the biological cell cycle with low potentials more easily seen at the bottom contour map.

In our earlier study of budding yeast system without explicitly putting in the excitations from $G1$ ground state to the start signal\cite{Wang2007BJ, Wang2008PRE}, we see quite different dynamics and landscape. There the potential landscape has a funneled shape. The system has one dominant basin of attraction pointing towards $G1$. The model can explain the dynamical process of yeast cell cycle once the start signal kicks off. Since the end state is always the $G1$ state (bottom of the funnel or basin of attraction), it does not contain the cycle part. The excitation from G1 state here giving the cycle is triggered by the nutrition supply or energy pump \cite{Wang2008PNAS, Xu2012, WangLiWangPNAS2010, LiWangPNAS2014}.

 With the excitation explicitly implemented in the fission yeast cell cycle, we can see in Fig.~\ref{fig:landscape} the potential landscape changes the shape from single attractor funnel to Mexican hat shape which can guarantee the stabilities of the oscillation states. If we see Mexican hat landscape a global quantification of a closed loop line attractor, an effective ``funneled'' potential landscape towards the oscillation path emerges. The degree of the funneless is quantified by RR. The ``funneled'' potential landscape towards the oscillation guarantees the global stabilities of the states on oscillation paths. However, this potential landscape can not guarantee the stable directional flows for the oscillations. The state  transitions  or switching along the biological path from the energy pump provided by the nutrition supply directs the oscillation cycle of the fission yeast cycle (G1 $\rightarrow$ S  $\rightarrow$ G2  $\rightarrow$ M  $\rightarrow$ G1). This is globally manifested by the underlying flux landscape, as we will show later in this study.

\subsubsection*{Robustness of potential landscape against changes in sharpness of response, self degradation, and stimulations.}

To further explore the robustness of the networks with internal and external perturbations, we calculate the Robustness Ratio (RR), probability of the cell cycle path, entropy production against the variations of different parameters.

\begin{figure*}[htp]
\includegraphics[width=0.85\textwidth]{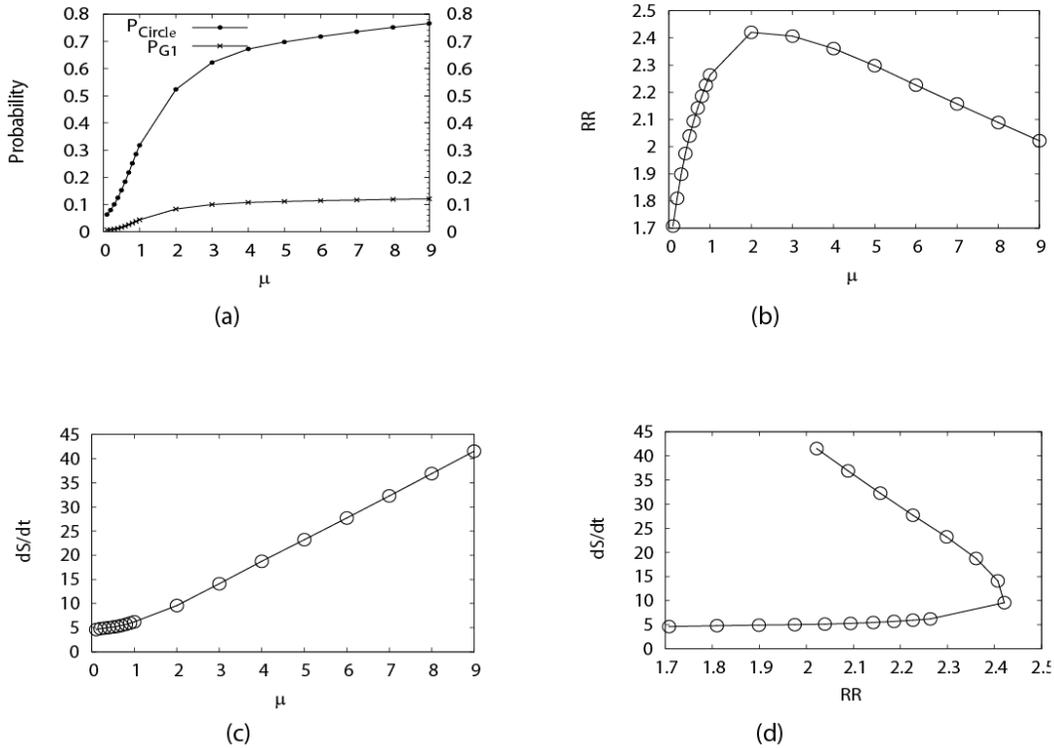}
\caption{Influence on the system robustness from the variation of the sharpness of the response or the inverse noise level $\mu$,  by fixing $c = 0.001$ and $\gamma=60\%$. (a) Steady-state probability of stationary G1 ($P_{G1}$) and "native" cycle ($P_{Cirle}$) versus $\mu$. (b) Robustness Ratio (RR) versus $\mu$. (c) Entropy production rate (${dS}/{dt}$) versus $\mu$. (d) Entropy production rate (${dS}/{dt}$) versus Robustness Ratio.}\label{fig:ppm}
\end{figure*}

In Fig.~\ref{fig:ppm}(a), we have shown the stationary probabilities for both $G1$ state and cell cycle path by fixing $c = 0.001$, $\gamma=60\%$ and changing $\mu$. It shows while the transition from input to output $\mu$ increases or the fluctuation decreases, the whole biological cell cycle, as well as the $G1$ state,  becomes more stable monotonously. When comparing the biological cell cycle with the $G1$ phase by changing $\mu$,  the biological cell cycle is more statable than the $G1$ phase with higher probabilities.

The changing of RR while switching the parameter $\mu$ is shown in Fig.~\ref{fig:ppm}(b). As we see at first the Robustness Ratio increases when $\mu$ increases or the fluctuation decreases.  This means a large transition from input to output or smaller fluctuations makes a more robust network. Notice that the RR goes down a little bit when $\mu$ is large enough. This result is similar to the behavior shown in the budding yeast cell cycle without the excitation from $G1$ ground state\cite{Wang2007BJ}. As the value $\mu$  is related to the inverse strength of the noise level, it can  be considered as the inverse temperature \cite{Wang2007BJ,Tang06}. Certain traps may appear which have low potentials and high weights at "low temperatures" (small fluctuations or high transition rate $\mu$). The presence of these traps leads the potential landscape spectrum less separated from the lowest potential state, and therefore less RR. One needs to increase the "temperature" with more fluctuations or decrease transition (or switching) $\mu$ (at high $\mu$) in order to "kick" the system out of the traps and therefore increase RR.

In Fig.~\ref{fig:ppm}(c) and Fig.~\ref{fig:ppm}(d), we plotted the entropy production rate or the dissipation cost of the network, $\frac{dS}{dt}$, against $\mu$ and RR.  We can see that the sharper the switching is, and therefore the more stable the oscillation is, the more dissipation cost is. The stable oscillation requires more energy consumptions to maintain it. The entropy production rate is the accumulated effects from the combination of both landscape and flux. Therefore the entropy production rate is in general a nonlinear function of these accumulated effects of landscape and flux. In addition, we see that traps consumes more energy and this leads to less stability of oscillation states through RR.

\begin{figure*}[htp]
{\includegraphics[width=0.85\textwidth]{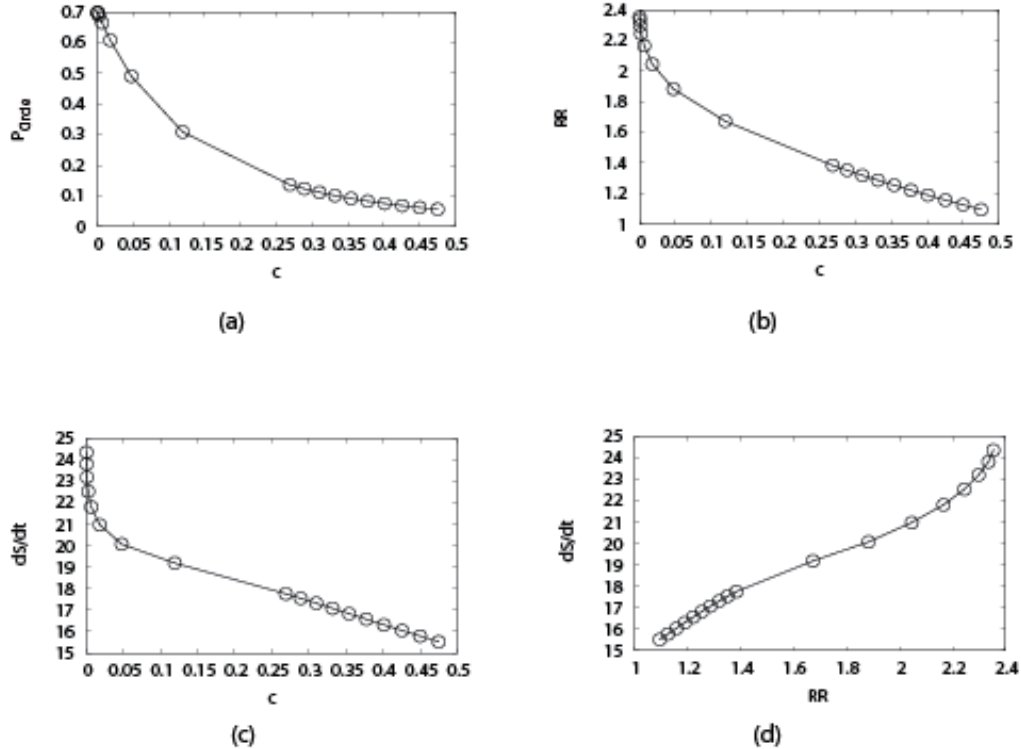}
\caption{Influence on the system robustness from the variation of the self-degradation parameters $c$,  by fixing $\mu=5, \gamma=60\%$. (a) Steady-state probability of "native" cycle ($P_{Cirle}$) versus $c$. (b) Robustness Ratio (RR) versus $c$. (c) Entropy production rate (${dS}/{dt}$) versus $c$. (d) Entropy production rate (${dS}/{dt}$) versus Robustness Ratio.}\label{fig:pc}}
\end{figure*}

Fig.~\ref{fig:pc}(a) shows the steady state probabilities of the biological cell cycle versus different self-degradation parameters $c$ (fixing $\mu=5,\ \gamma=60\%$). We notice that large (small) $c$ indicates large (small) self-degradation, and then the probability of biological cell cycle is decreasing with respect to $c$. It indicates that less self-degradation gives more stable biological cell cycle, and therefore a more robust network which is shown in Fig.~\ref{fig:pc}(b). Fig.~\ref{fig:pc}(c) and Fig.~\ref{fig:pc}(d) show the dissipation decreases as the self-degradation increases and as the RR decreases, which shows that more stable system needs more energy consumption to maintain it.

\begin{figure*}[htp]
{\includegraphics[width=0.85\textwidth]{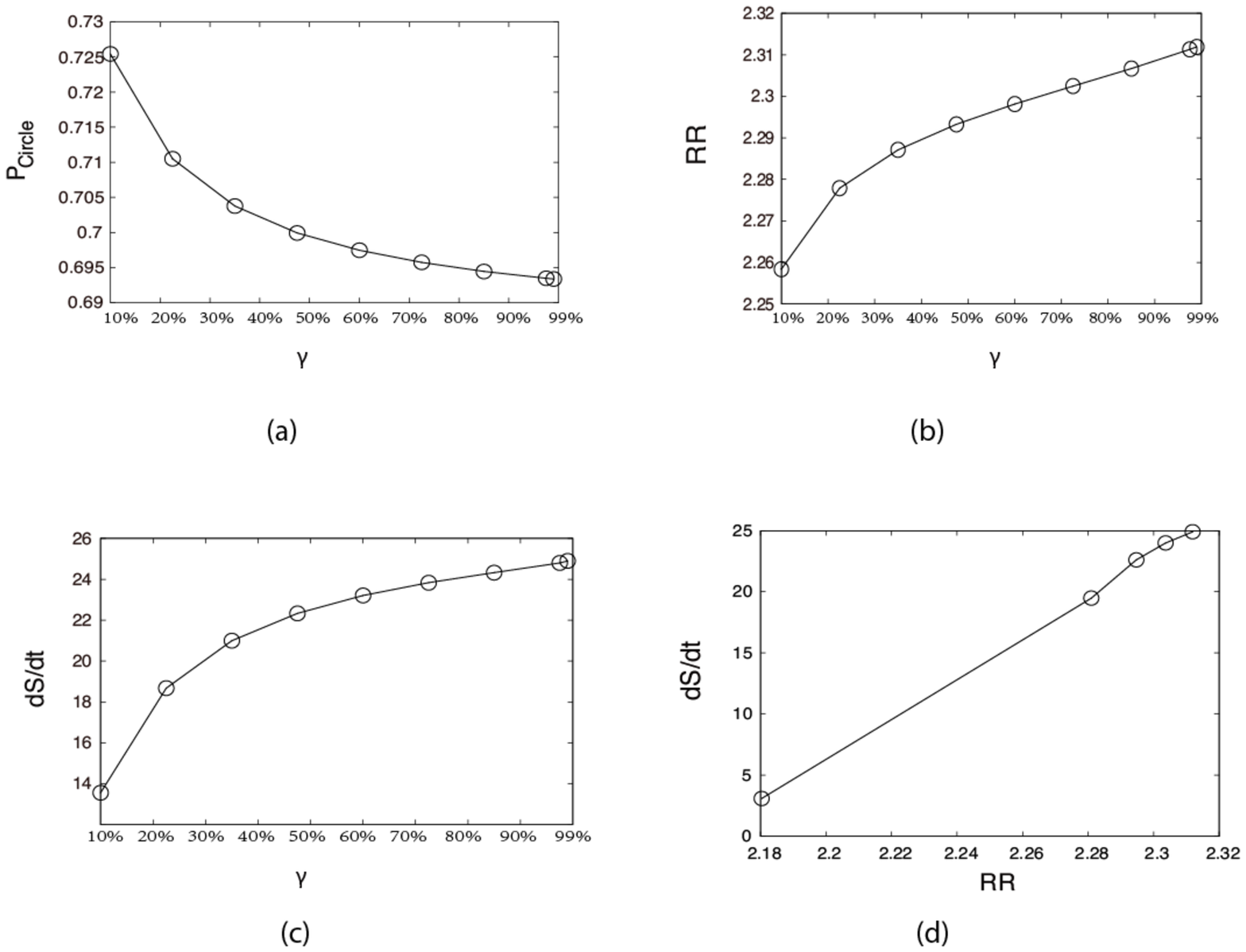}
\caption{Influence on the system robustness by changing the stimulation  $\gamma$ which represents the jumping probability from the checkpoint G1 to the Start state,  while fixing $\mu=5$, $c = 0.001$. (a) Steady-state probability of "native" cycle ($P_{Cirle}$) versus $\gamma$. (b) Robustness Ratio (RR) versus $\gamma$. (c) Entropy production rate (${dS}/{dt}$) versus $\gamma$. (d) Entropy production rate (${dS}/{dt}$) versus Robustness Ratio.}\label{fig:pg}}
\end{figure*}

$\gamma$ can be considered as the jumping probability which represents that the state of the fission yeast receives a start signal to begin from the $G1$ phase.  We see in Fig.~\ref{fig:pg}(a) that the weights or occupational probabilities of the states on the oscillation path do not change significantly with respect to $\gamma$. Although $\gamma$ does not change the stabilities of these oscillating states much, it does lead to the directed flow and therefore the stable oscillations as seen in the later flux landscape discussions. Consequently, stimulations through $\gamma$ does not change significantly the shape of the potential landscape as shown in Fig.~\ref{fig:pg}(b). Fig.~\ref{fig:pg}(c) and Fig.~\ref{fig:pg}(d) show that the recycling probability of the cell cycle costs more energy and maintaining a more stable oscillating system requires more energy consumption.

\begin{figure*}[htp]
{\includegraphics[width=0.85\textwidth]{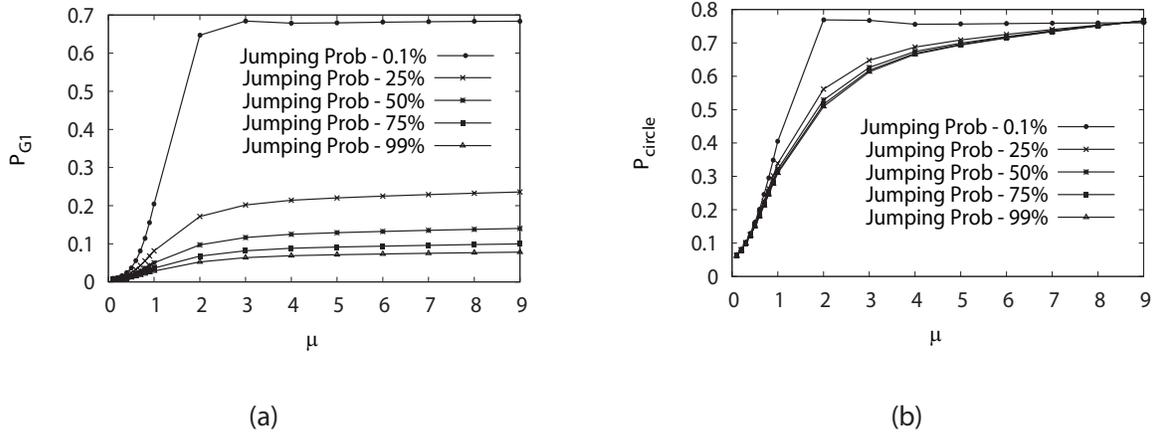}
\caption{Steady-state probability of stationary G1 ($P_{G1}$) (a) and probability of "native" cycle ($P_{Cirle}$) (b) versus the variation of $\mu$ under different jumping probabilities $\gamma$ by fixing $c = 0.001$.}\label{fig:pggm}}
\end{figure*}

Fig.~\ref{fig:pggm}(a) and Fig.~\ref{fig:pggm}(b) show the steady state probability of $G1$ state and the steady state probability of the biological cell cycle path under various stimulation levels ($\gamma$) at different switching or fluctuations  $\mu$. It has been studied above that $\mu$ should characterize the inverse  noise strength and increase of which enhances the stability of both the $G1$ phase and the biological cell cycle path. When considering the changes of jumping probability $\gamma$, one can see in Fig.~\ref{fig:pggm}(a) and Fig.~\ref{fig:pggm}(b) that the larger the stimulation probability from $G1$ phase is,the less stable $G1$ state becomes. The weights or the occupation probabilities of the states on the oscillation paths are also decreased slightly with respect to stimulation $\gamma$, but not significantly compared to that of G1. This means that although the individual states or phases of the cell cycle such as G1 becomes less stable, the stimulation does not change significantly the overall occupations of the states on the oscillation paths and therefore the associated stability.

\subsubsection*{Power spectrum under different stimulation and sharpness of responses}
To illustrate the oscillation cycle, one can introduce the correlation functions of the dynamical observables. For the fission yeast cell cycle, the observables are the expressions of individual genes. This gives the measure on the dynamical response and fluctuations of the systems. The Fourier transform of the autocorrelation function in time of gene expression variables gives the power spectrum.  The power spectrum analysis is widely used in the periodical signal transduction mixed with noise interference \cite{Norton2003}. The gene expression time series of fission yeast cell cycle can be viewed as the noisy signals.  We try to calculate the power spectrum of the fission yeast cell cycle to find how oscillation is influenced by different parameters, and then study their relationship with the flux landscape in the global state space.

\begin{figure*}[htp]
{\includegraphics[width=0.85\textwidth]{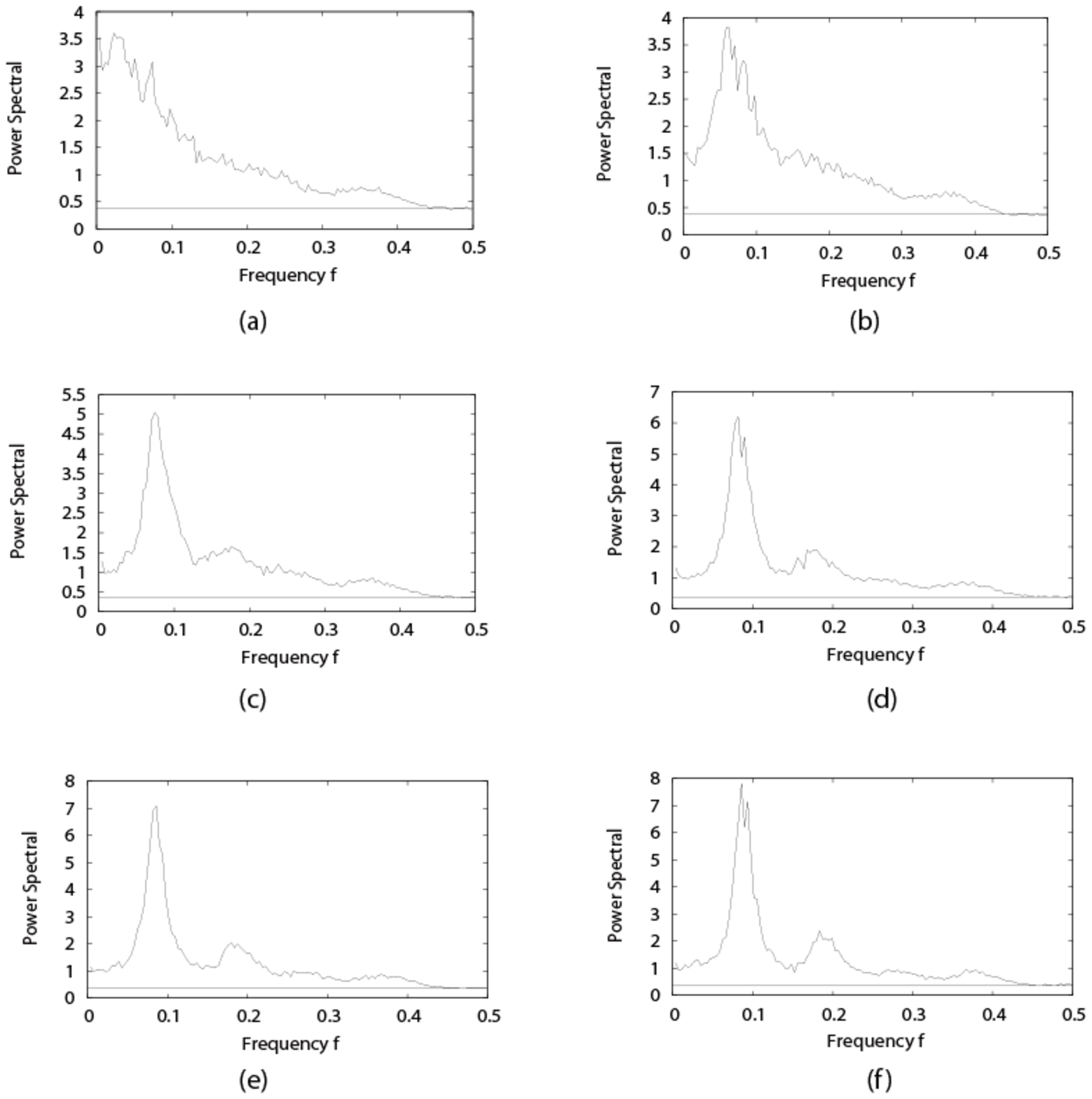}}
\caption{Power spectrum of the fission yeast cell cycle by changing the jumping probability $\gamma=\{1\%,11\%,31\%,51\%,71\%,99\%\}$ for (a)-(f), while fixing $c = 0.001, \mu=5$.}\label{fig:powc01}
\end{figure*}

The results from Fig.\ref{fig:powc01}(a) to Fig.\ref{fig:powc01}(f) show that a main peak of the power spectrum emerges and becomes more prominent with the jumping probability $\gamma$ being larger ($0<\gamma<0.4$) when fixing $c = 0.001, \mu=5$. This peak corresponds to exactly the speed or frequency of the cell cycle oscillation path. This shows that the intrinsic frequency of the oscillations from the loop flux (the loop flux is proportional to the speed and inversely related to the period of oscillations) coincides with the external response frequency at power spectrum peak.  In other words, this results the resonance from the system's external response by power spectrum to reflect the intrinsic frequency of the cell cycle oscillations. When $\gamma$ becomes larger than $0.4$, we also find that the power spectrum  will have even more significate peaks. As one peak can map into one periodical oscillation, we can see that with jumping probability $\gamma$  becoming higher, there are more and more smaller peaks representing smaller loops emerging, and the biological cell cycle path tends towards the most prominent loop, which can be considered as the dominant loop compared with others.

\begin{figure*}[htp]
  \centering
{\includegraphics[width=0.85\textwidth]{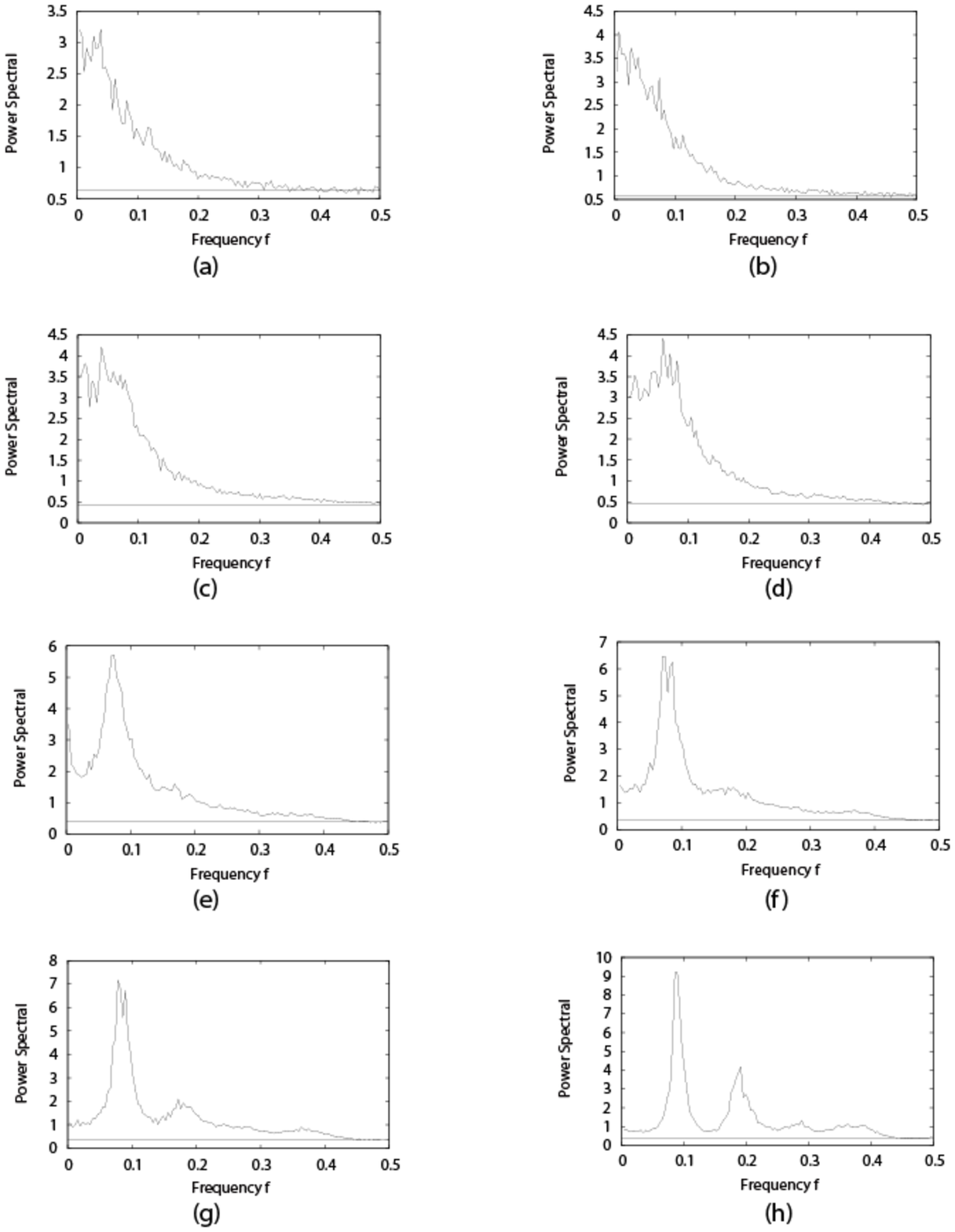}}
\caption{Power spectrum of the fission yeast cell cycle by changing $\mu=\{0.1,0.5,0.9,1.0,2.0,3.0,5.0,9.0\}$ for (a)-(h), with $c = 0.001, \gamma=60\%$.}\label{fig:powb01}
\end{figure*}

While by setting $c = 0.001$, $\gamma=60\%$ with changing $\mu$, Fig.\ref{fig:powb01}(a) to Fig.\ref{fig:powb01}(h) give us a clear view that the increasing $\mu$ leads to one prominent peak and several other smaller peaks, which again shows that the fission yeast cell cycle contains one main dominant periodical oscillation, and many other much smaller cycles. At even larger stimulations chaos might emerge.

\begin{figure*}[htp]
\centering
\includegraphics[width=0.85\textwidth,angle=-90]{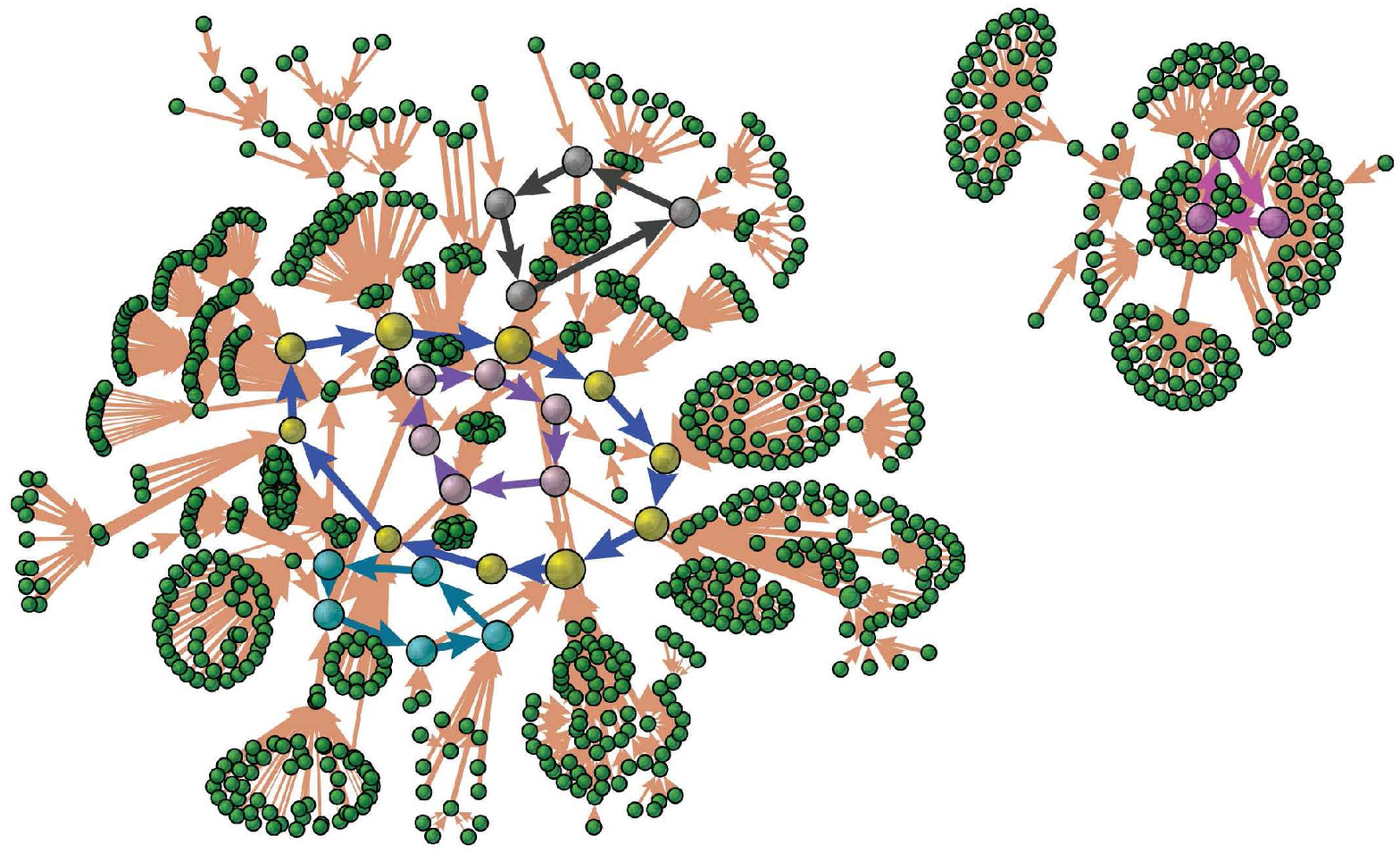}
\caption{The probability flux flowing in the $2^{10}$ states space of the fission yeast cell cycle. The thickness of the arrows is proportional to the magnitude of the probability fluxes and the node size is related to the steady state probability of the state. The colored flux loops represent several typical flux loops among all, in which the largest blue loop with $10$ nodes stands out and becomes the "native" biological cycle. The simulations were performed with $c = 0.001, \mu=5, \gamma=60\%$.}\label{fig:network}
\end{figure*}

\subsection*{Funneled flux landscape leads to robust limit cycle oscillations}
\subsubsection*{Quantifying the flux landscape}

To give a whole picture of the probabilistic flux, we further calculate the flux landscape of this fission yeast cell cycle, and analyze how the flux landscape is influenced by different parameter variations, such as $\mu$, $c$ and $\gamma$. By comparing the characteristics of landscape of flux with that of the landscape of potential, we expect to gain unique insights on the non-equilibrium biological cell cycle.

\begin{figure}[htp]
\includegraphics[width=0.450\textwidth]{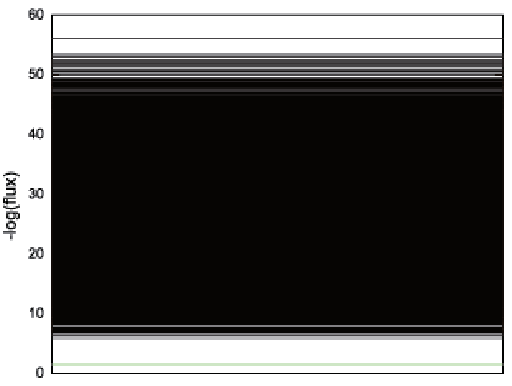}
\caption{Probability flux spectrum of all the loops, where $\mu = 5$, $c = 0.001$ and
 $\gamma=60\%$. The flux of the loop which is formed by the 10 states of the biological pathway and thus represents the "native" cycle is drawn in green line. It is the lowest one in negative logarithm compared to other cycle loops.}\label{fig:fspect}
\end{figure}

{After the decomposition of the driving force into the probability landscape and probability flux, we further decompose the probability flux into the cycle loops. This forms the flux landscape with different loops. The results are shown in Fig.\ref{fig:network} (we did not show all the loops for the purpose of clear view on the figure). The thickness of the arrows represents the magnitude of the probability fluxes and the node size represents the steady state probability of that state. We can see that the blue loop can be considered as a dominant loop of the biological cell cycle path with dominant flux flowing along the $10$ states of the cell cycle. There are also other secondary loops which can map into those secondary peaks in Fig.\ref{fig:powc01}(f) or Fig.\ref{fig:powb01}(c). We notice that there is a small number of states clustered together but separated from the major state cluster. These are possible traps of states.

To further quantifying the flux landscape, we show that the flux landscape spectrum in Fig. \ref{fig:fspect}. We can see that there is a clear separation seen from the large gap between the flux from "native" cycle and the rest of the others under the chosen parameter set. This means the cell cycle loop stands out from the sea of many loops and becomes dominant. As a result, the flux landscape is funneled towards the dominant loop state. This provides a physical picture for the origin of the limit cycle.}

\subsubsection*{Robustness of flux landscapes against changes in sharpness of the response, self degradations and stimulations}

\begin{figure*}[htp]
  {\includegraphics[width=0.85\textwidth]{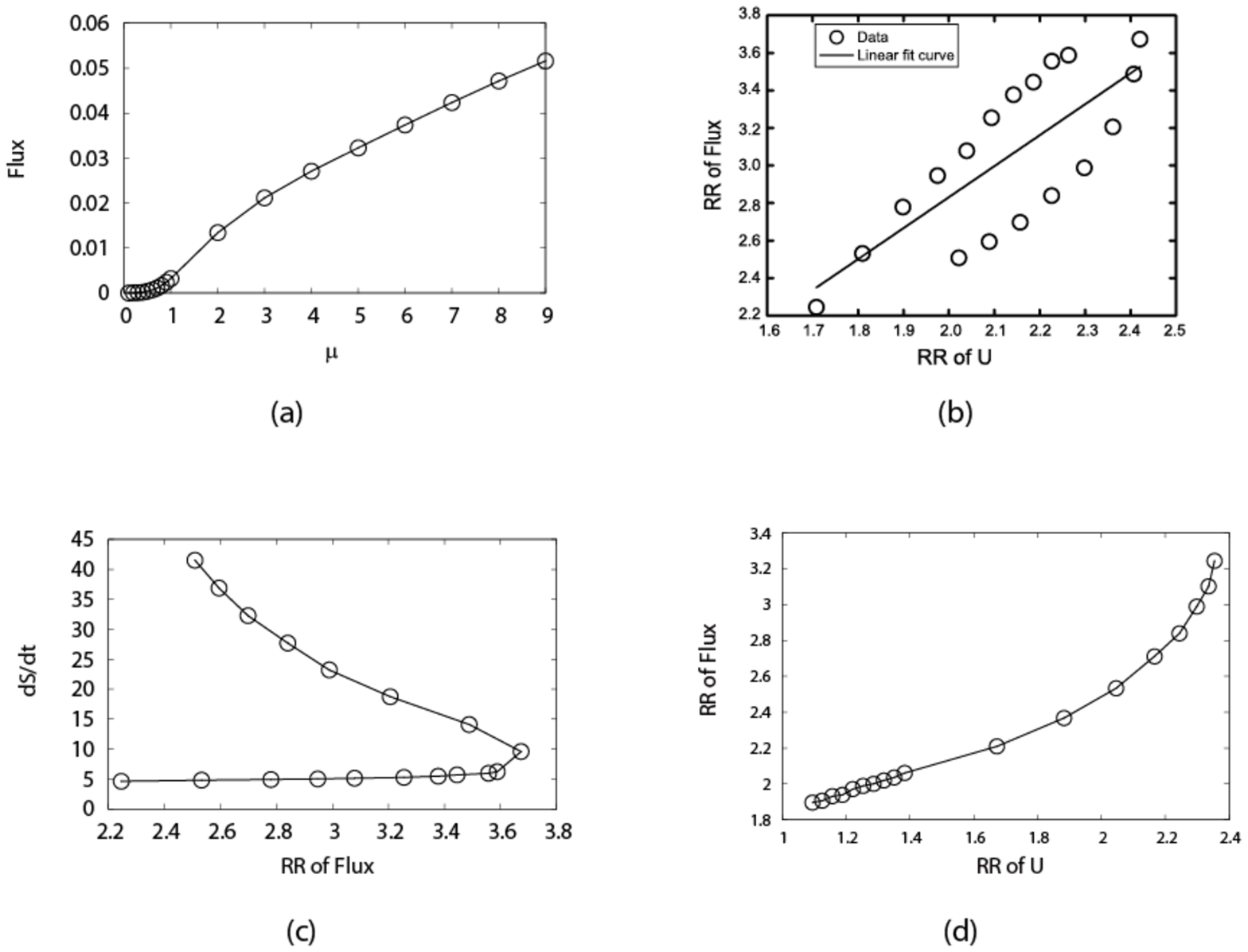}}
\caption{Influence on the probability flux from the variation of the sharpness of the response or the inverse noise level $\mu$,  by fixing $c = 0.001$ and $\gamma=60\%$. (a) Steady-state probability flux versus $\mu$. (b) Robustness Ratio (RR) of flux spectrum versus $\mu$. (c) Entropy production rate (${dS}/{dt}$) versus RR of flux. (d) Robustness Ratio (RR) of flux spectrum  versus RR of potential landscape spectrum. }\label{fig:ppmf}
\end{figure*}

We plot the dominant flux flowing along the biological cell cycle, by varying response $\mu$ and fixing parameters self degradation $c = 0.001$ and stimulation $\gamma=60\%$ in Fig.~\ref{fig:ppmf}(a). We can see that the increasing  $\mu$ can directly lead to the increase of the flux. Reminding the discussions above on the robustness of potential landscape (Fig.~\ref{fig:ppm}(a)), we can state that high probability of switching between states or less environment fluctuations will lead to higher flux of the cell cycle.

We explore the robustness RR of the flux landscape in Fig.~\ref{fig:ppmf}(b) and Fig.~\ref{fig:ppmf}(c) to show how it relates to  $\mu$ and entropy production rate $dS/dt$. As we have mentioned, increasing $\mu$ can lead to higher probability of the G1 ground state, as well as the cycle flux. Here we see the robustness RR also increases to certain level. The excessively large $\mu$ can lead to higher probabilities of other states as traps near G1. This results to the decrease of RR of both potential landscape and flux landscape. Larger fluctuations (smaller $\mu$) from the large $\mu$ side will lead to the "escape" from the traps and therefore larger RR as shown in Fig.~\ref{fig:ppmf}(b).  Fig.~\ref{fig:ppmf}(c)  indicates that more robust oscillating cell cycle often costs more energy to maintain, while traps can consume even more energies. Comparing with the potential landscape shown in Fig.~\ref{fig:ppm}(b) and Fig.~\ref{fig:ppm}(d), we obtain Fig.~\ref{fig:ppmf}(d), which shows the consistency of the robustness measure of the potential landscape using states on cell cycle as "native" states and the robustness measure of the flux landscape.

\begin{figure*}[htp]
 {\includegraphics[width=0.85\textwidth]{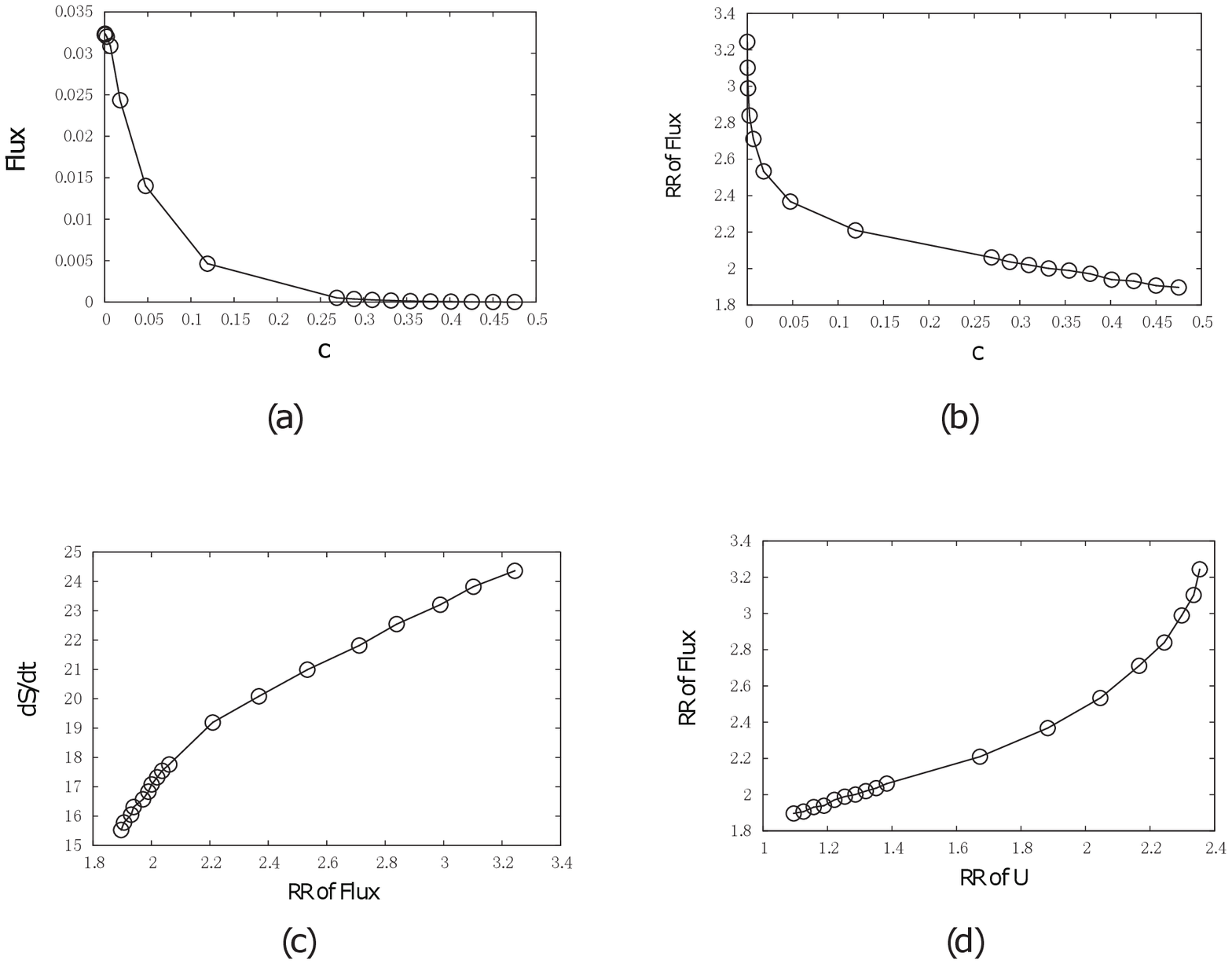}}
\caption{Influence on the probability flux  from the variation of the self-degradation parameters $c$,  by fixing $\mu=5, \gamma=60\%$. (a)  Steady-state probability flux versus $c$. (b) Robustness Ratio (RR) of flux spectrum versus $c$. (c) Entropy production rate (${dS}/{dt}$) versus  RR of flux. (d) Robustness Ratio (RR) of flux spectrum  versus RR of potential landscape spectrum.}\label{fig:pcf}
\end{figure*}

We also calculate how flux landscape is influenced by the self-degradation parameter $c$ (by fixing $\mu=5, \gamma=60\%$). From Fig.~\ref{fig:pcf}(a)-Fig.~\ref{fig:pcf}(c), when the self-degradation $c$ decreases, the cycle flux increases and the funneled flux landscape becomes more robust, which costs more energy to maintain. Fig.~\ref{fig:pcf}(d) also shows that the robustness of the flux landscape is monotonic related to the robustness of the potential landscape upon varying of $c$.

\begin{figure*}[htp]
  {\includegraphics[width=0.85\textwidth]{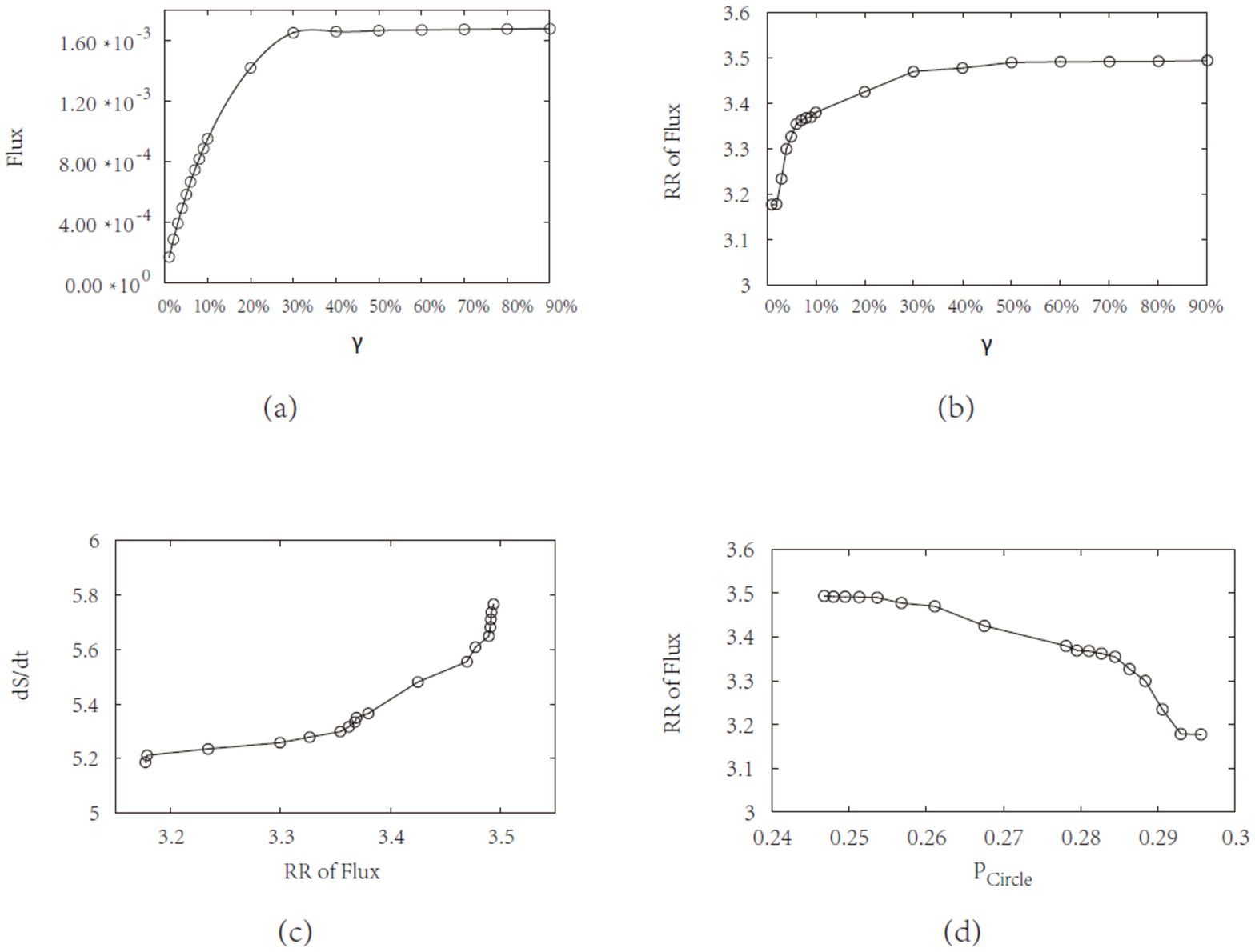}}
\caption{Influence on the probability flux by changing $\gamma$ which represents the jumping probability from the checkpoint G1 to the Start state,  while fixing $\mu=0.8$, $c = 0.001$. (a) Steady-state probability flux versus $\gamma$. (b) Robustness Ratio (RR) of flux spectrum versus $\gamma$. (c) Entropy production rate (${dS}/{dt}$) versus  RR of flux. (d) Robustness Ratio (RR) of flux spectrum versus steady-state probability of "native" cycle ($P_{Cirle}$).}\label{fig:pgf}
\end{figure*}

We study the relationship between the stimulation or pumping strength from the G1 state to start state and the flux landscape. The flux originates from the the nutrition supply which provides the energy pump (for example, thorough the release of the ATP production). We show in Fig.~\ref{fig:pgf}(a)-Fig.~\ref{fig:pgf}(d) how the pumping strength $\gamma$ influences the shape or topography of the underlying flux landscape. Before the pumping, the biological cell cycle is an one-way stable pathway to G1\cite{Tang04,Wang2007BJ}. At this stage, even the occupations of the states of the oscillation path including G1 are higher with high RR for potential landscape, the robust directional flow along the cycle has not been formed yet. As a result, no oscillations are emerged.  The increase of pumping strength can be considered as a switch to form the flux landscape with cycle loops from a deep stable one basin potential landscape. When the pumping increases, many cycle loops start to form and flux landscape starts to emerge. When pumping $\gamma$ is large enough the flux and RR reaches saturation. Further pumping will not be effective since robust cycle and flux landscape has already been formed. The average cycle flux increases and the robustness of the flux landscape increases. This indicates that a single flux loop dominates and stands out from the rest. It leads to robust cell cycle. The robust cycle costs more energy to maintain.

Fig.~\ref{fig:pgf}(d) shows that although the increase of the stimulation or pumping from the G1 to the start of the cell cycle leads to a slight decrease of the occupations of the states on the oscillating path (due to the excitations), the separation between the dominant flux loop and the rest of the decoys increases. As a result, the dominant flux loop stands out and forms the directional flow along the yeast cell cycle.

\begin{figure*}[htp]
  {\includegraphics[width=0.85\textwidth]{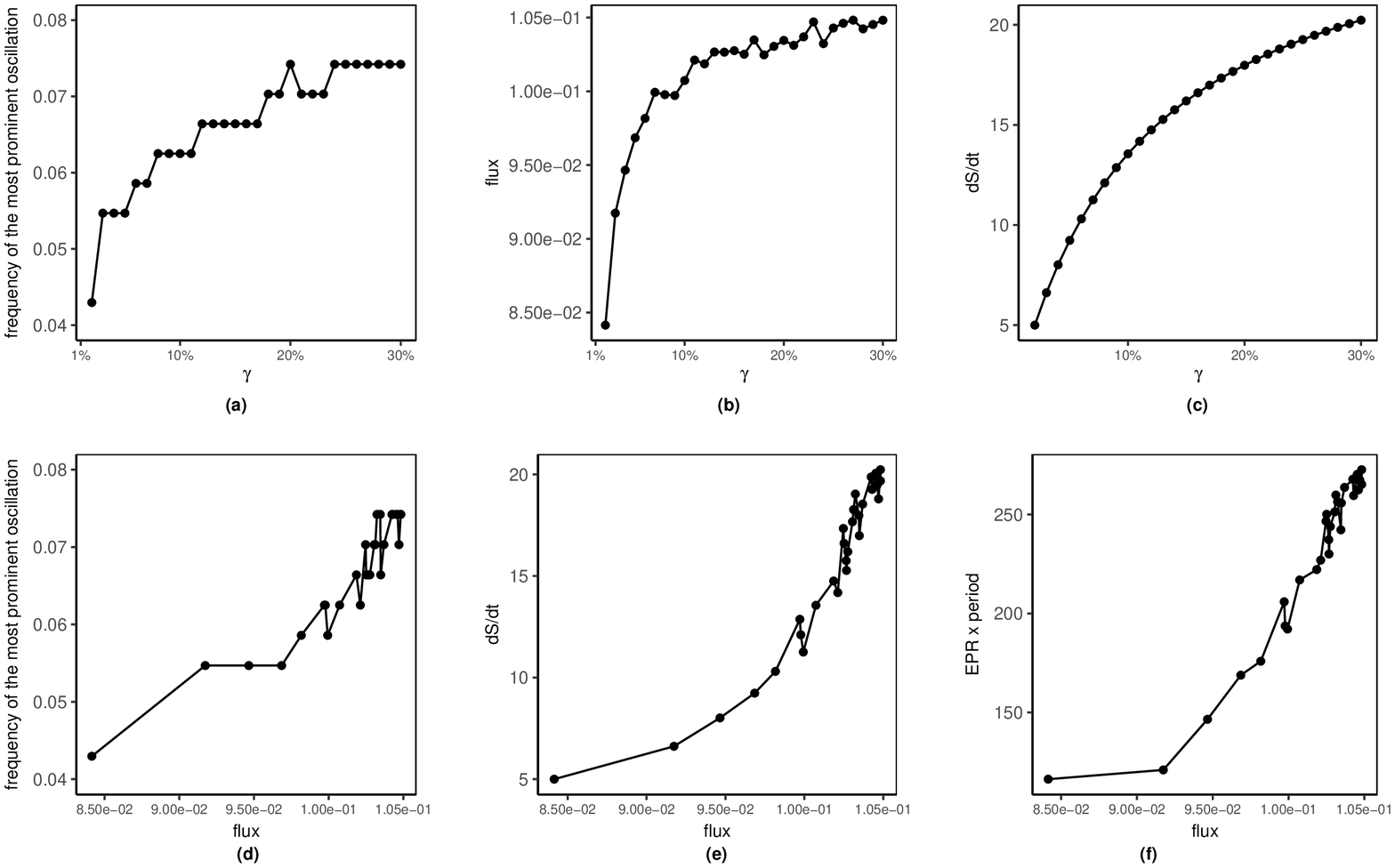}}
\caption{Relationship between the most prominent frequency, entropy production and flux by changing $\gamma=(1\%,30\%)$, while fixing $\mu=5$, $c = 0.001$. (a)-(c) Variation of most prominent frequency, flux and entropy production rate when changing the pumping strength $\gamma$. (d)-(f) Positive correlation between the most prominent frequency, entropy production and flux when changing $\gamma$. }\label{fig:gcf}
\end{figure*}
\subsubsection*{Energy pump, curl flux, dissipation, speed of the cell cycle, and origin of life}

To further explore the intrinsic mechanism of flux landscape formation, we study the relationship between the oscillation speed, entropy production and flux.  From Fig.~\ref{fig:powc01}, we can find that when the pumping strength $\gamma$ increase from 1\% to 30\%, the frequency of most prominent oscillation also increases. We can associate the most prominent frequency and flux.  We focus on these relationships in Fig.~\ref{fig:pgf}(a),  (b)and (d), in which we can see clearly that larger flux leads to the faster cell cycle oscillation. Therefore, the energy pump through $\gamma$ is the origin of the flux and  the flux drives the limit cycle and determines the associated speed or period.

Furthermore, we also explore the relationship between entropy production and flux under
pumping strength $\gamma$ changes in Fig.~\ref{fig:gcf}(c), (e)and (f).  The entropy production rate here represents the energy dissipation in steady state. These figures show that larger flux gives larger energy dissipation and the total entropy production becomes larger. This demonstrates that the degree of detailed balance breaking from the energy pump measured by the flux is the cause of the energy dissipation for sustaining the oscillation.

A fundamental signature of living is the biological replications which can be described by the limit cycle oscillations. From this quantitative study, we can see that energy pump is required for the replications to emerge and survive. Therefore energy supply is necessary for life. This also has evolution implications for the origin of life being initiated and sustained by energy supply.

As a summary, we can state that the limit cycle oscillation is maintained to be stable due to two driving forces: the funneled potential landscape which tends to attract the system down to the close ring valley, leading to high occupations of the states along the oscillation path. The directional flow along the oscillation path is driven by the probability flux originated from the nutrition supply manifested as the stimulation or excitation from G1 to the start of cell cycle. The funneled flux landscape guarantees the clear separation between dominant flux loop and the rest of the other flux loops. Consequently, the dominant flux loop stands out and forms the yeast cell (limit) cycle.   Both forces from potential landscape and flux landscape are essential for the stability of the fission yeast cell cycle.

\subsection*{Backbone subnetwork of fission yeast from the evaluation of global stability \& robustness of both potential landscape and flux landscape}

We have uncovered that both the potential landscape and the flux landscape are crucial for the stability and the robustness of the limit cycle oscillation of the fission yeast cell cycle. As a practical application, we will perform global sensitivity analysis based on the two landscapes, to explore a backbone subnetwork to carry out
the biological functions.   We firstly perform perturbations through adding, deleting or repressing arrows between nodes in the wiring diagram in Fig.~\ref{fig:wiring}, or replacing an active arrow with an inactive arrow, or deleting an individual node. And then we try to analyze the variation of the important characteristic of the two landscapes, such as $RR$, $P_{G1}$, $P_{Cirle}$ $RR~of~flux$, and so on. Finally, we try to work out which key links or nodes are responsible for the stability, speed (function), and robustness of the cell cycle.

\subsubsection*{Global stability \& robustness of potential landscape under perturbations of mutations and regulation strengths}

\begin{figure*}[htp]
{\includegraphics[width=0.75\textwidth]{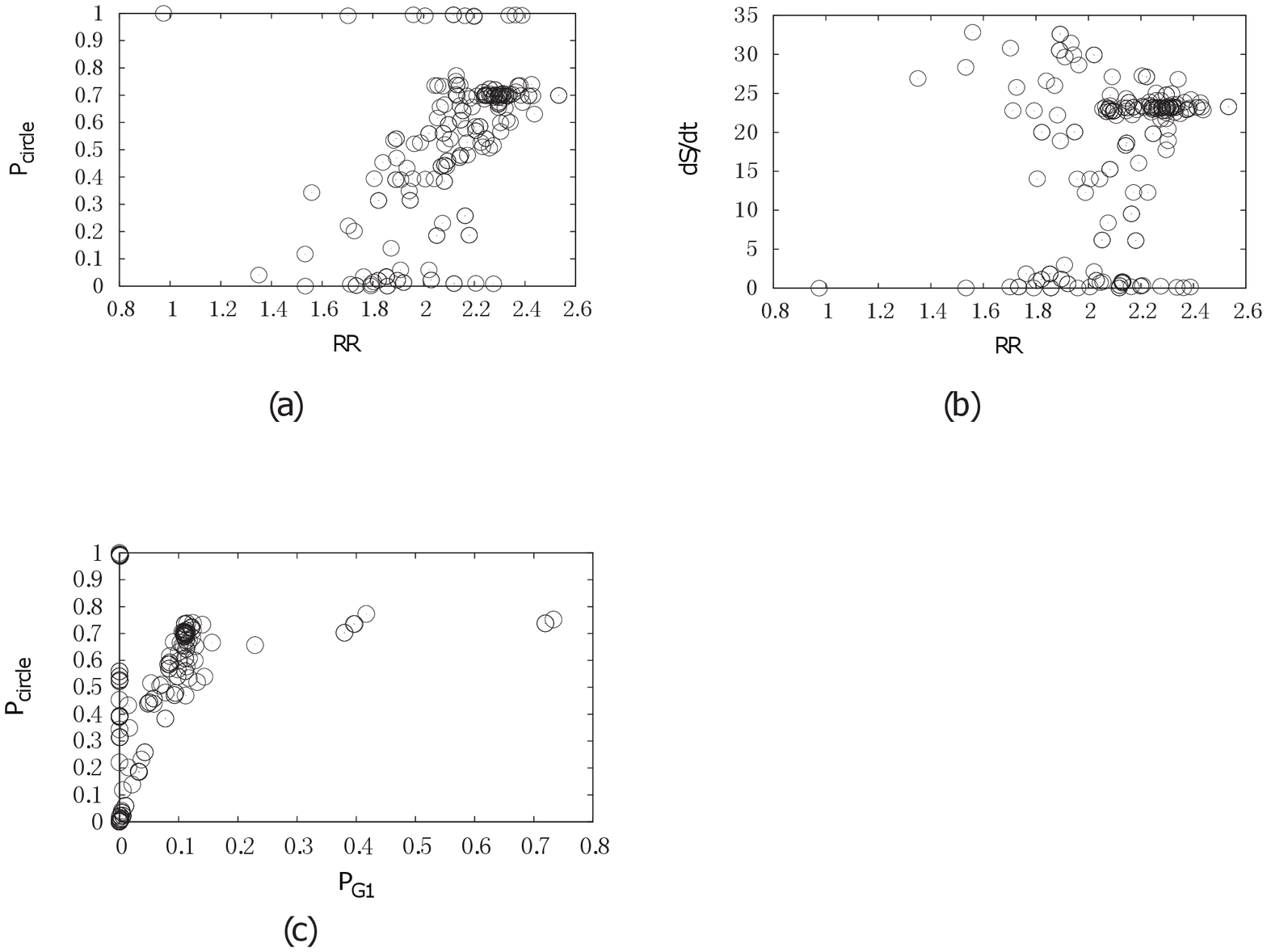}}
\caption{Perturbations through mutations (adding, deleting or repressing the arrows) on the fission yeast cell cycle network, while $\mu = 5$, $c = 0.001$ and $\gamma=60\%$. (a) Steady-state probability of "native" cycle ($P_{Cirle}$) versus Robustness Ratio under different perturbations. (b) Entropy production rate versus Robustness Ratio under different perturbations. (c) Steady-state probability of "native" cycle ($P_{Cirle}$) versus  steady-state probability of G1 ($P_{G1}$) under different perturbations.}\label{fig:perpr}
\end{figure*}

Fig.~\ref{fig:perpr}(a) shows the RR versus the probability of the biological cell cycle (fixing $\mu=5, \gamma=60\%$) against various perturbations. The perturbations are through adding, deleting or repressing the arrows between the nodes in the wiring diagram in Fig.~\ref{fig:wiring}, or replacing an activating arrow with an inactivating arrow, or deleting an individual node. We see the larger the RR is, the higher the occupation is of the states on oscillation cell cycle upon perturbations of links and nodes. This indicates the more stabilities of the states on the cell cycle. This provides the rational of using RR as a robustness measure for the cell cycle network. Fig.~\ref{fig:perpr}(b)  shows again that more stable oscillations requires more energy consumption. The points of low dissipations with large RR values are ignored, as the corresponding probabilities of the biological cycle path are low.

Fig.~\ref{fig:perpr}(c) shows the relationship between the steady state probability of the biological cell cycle path and the steady state probability of the $G1$ state upon perturbations of links and nodes. The stable $G1$ often correlates with higher occupations of the states on oscillating cycle paths, as $G1$ is the starting point in the biological cycle path. However the higher occupations of the biological cycle do not always imply stable $G1$ state, as quite a few perturbations will disable the stability of state $G1$ and enhance the stabilities of some other states along the cell cycle path.


\subsubsection*{{Global stabilities and robustness of flux landscape under perturbations of mutations and regulation strengths}}

\begin{figure}[htp]
\centering
{\includegraphics[width=0.45\textwidth]{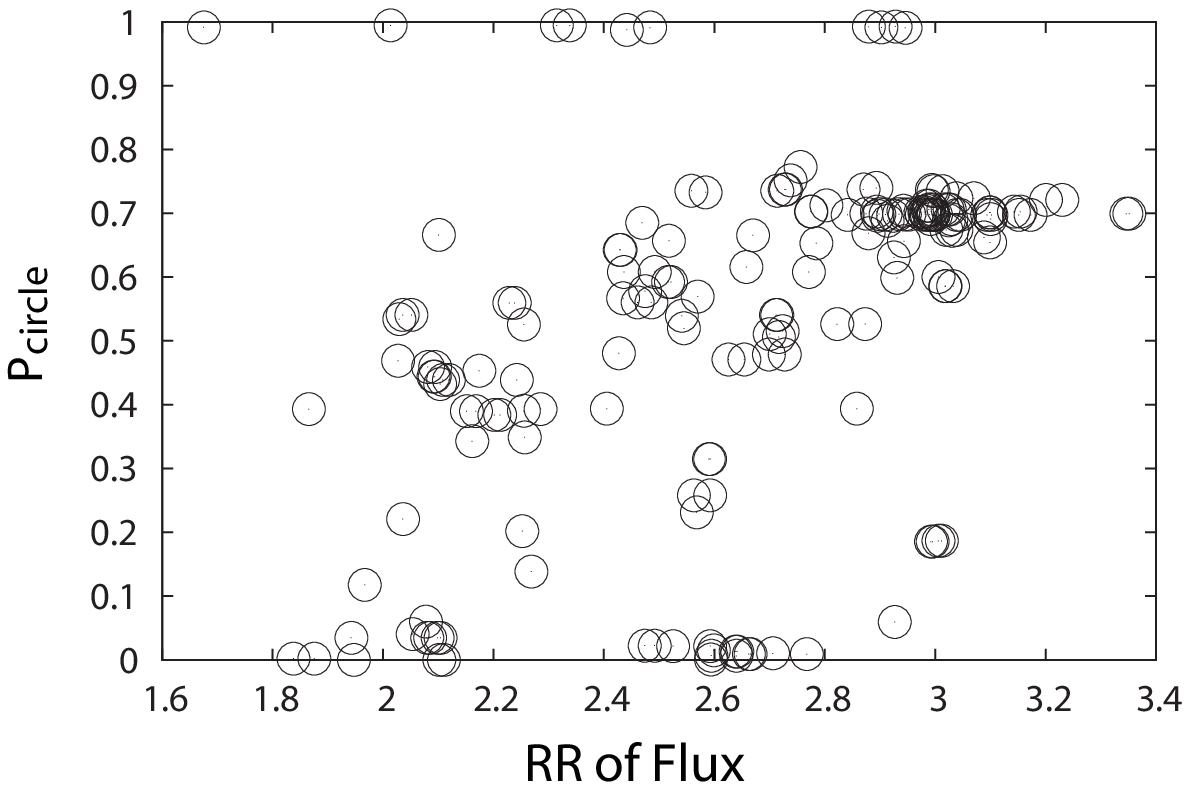}}
\caption{ Steady-state probability of "native" cycle ($P_{Cirle}$) versus RR of flux spectrum. It shows the perturbations through mutations (adding, deleting or repressing the arrows) on the fission yeast cell cycle network, while $\mu = 5$, $c = 0.001$ and $\gamma=60\%$.  }\label{fig:perprf}
\end{figure}

Fig.~\ref{fig:perprf}   shows the probabilities of oscillation cell cycle path and robustness ratios of the flux landscapes,  upon those mentioned perturbations.

We see that upon perturbations on links and nodes of the fission yeast cell cycle network, higher probabilities of states on oscillation path often accompany with higher robustness ratio of the flux landscape separating the cell cycle loop from the rest as shown in Fig.~\ref{fig:perprf}. There are some exceptions where very high and very low probabilities of the states on oscillation cell cycle path have varying robustness ratios of the flux landscapes. We can ignore those due to the insignificant cell cycle or insignificant decoys.

\subsection*{Backbone subnetwork contained in the fission yeast networks}

Through the global stability analysis for the key wirings of the networks upon perturbations of links and nodes, one can identify the key network structure elements or motifs responsible for the stability and biological function. To further identify the stable and functional backbone subnetwork, we choose to delete the link one by one and find out which will make the network more unstable. Based on the discussion in the text above, we select three essential elements to measure the importance of each network edge, that is $\bf RR$, $\bf P_{G1}$ (probability of the G1 steady state),  and $\bf RRflux$ (RR of flux landscape), which can  represent not only the global stability of G1 state but also the robustness of the biological path.

\begin{figure*}[h]
\centering
{\includegraphics[width=0.45\textwidth]{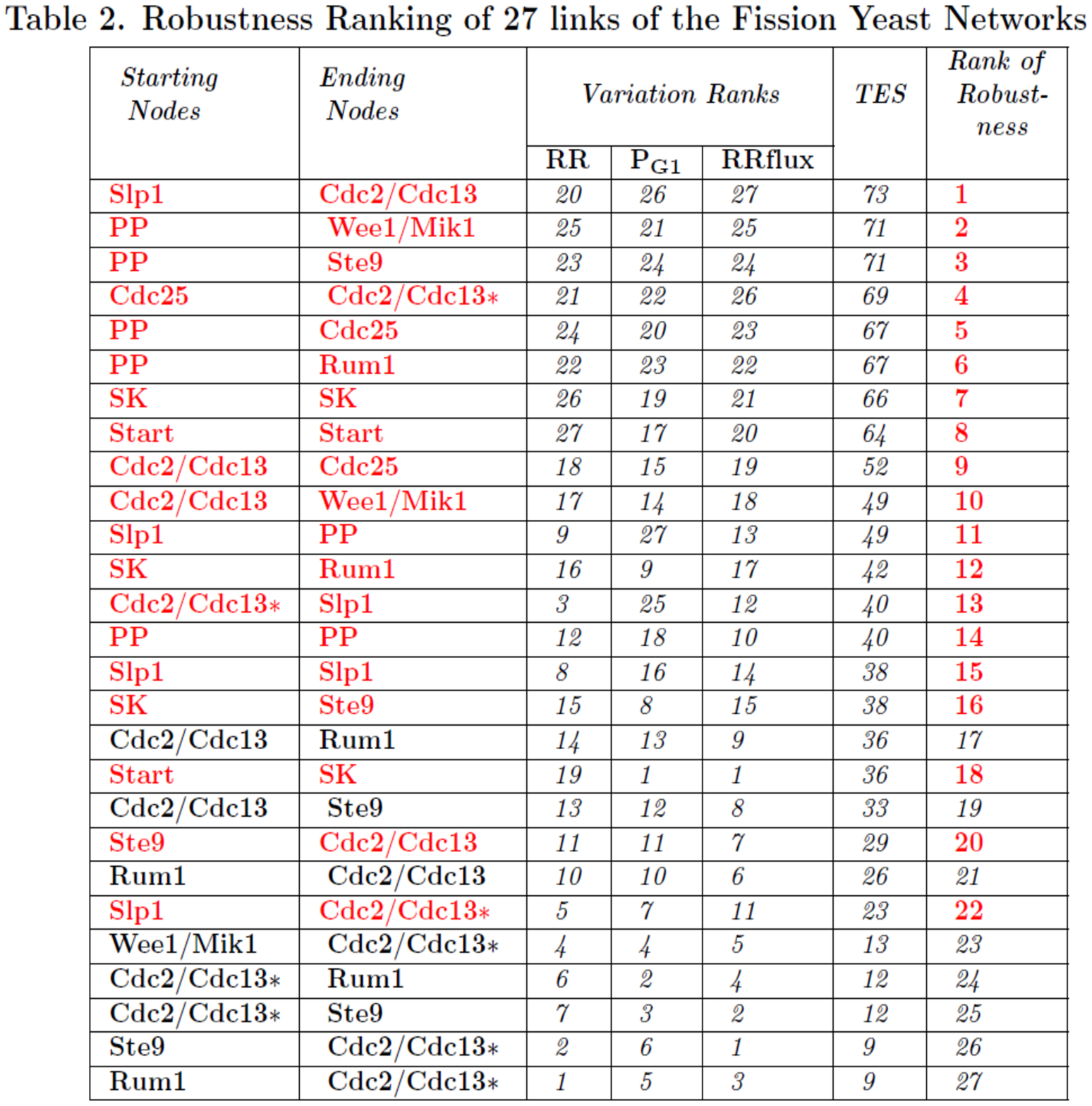}}
\label{table2}
\end{figure*}

The way we followed is like this: first of all, for each edge, we respectively calculating the difference of the three essential elements between mutated network and original network. Then we rank the difference of the 27 edges and give them score from 1 to 27. The larger is the difference, the larger the score is.  Then we rank the summation of all the three scores as a total evaluation score(TES), and obtain the rank of robustness of all the 27 edges of the fission yeast network. (One can see the result in Table.~\ref{table2})

\begin{figure*}[h]
\centering
{\includegraphics[width=0.45\textwidth]{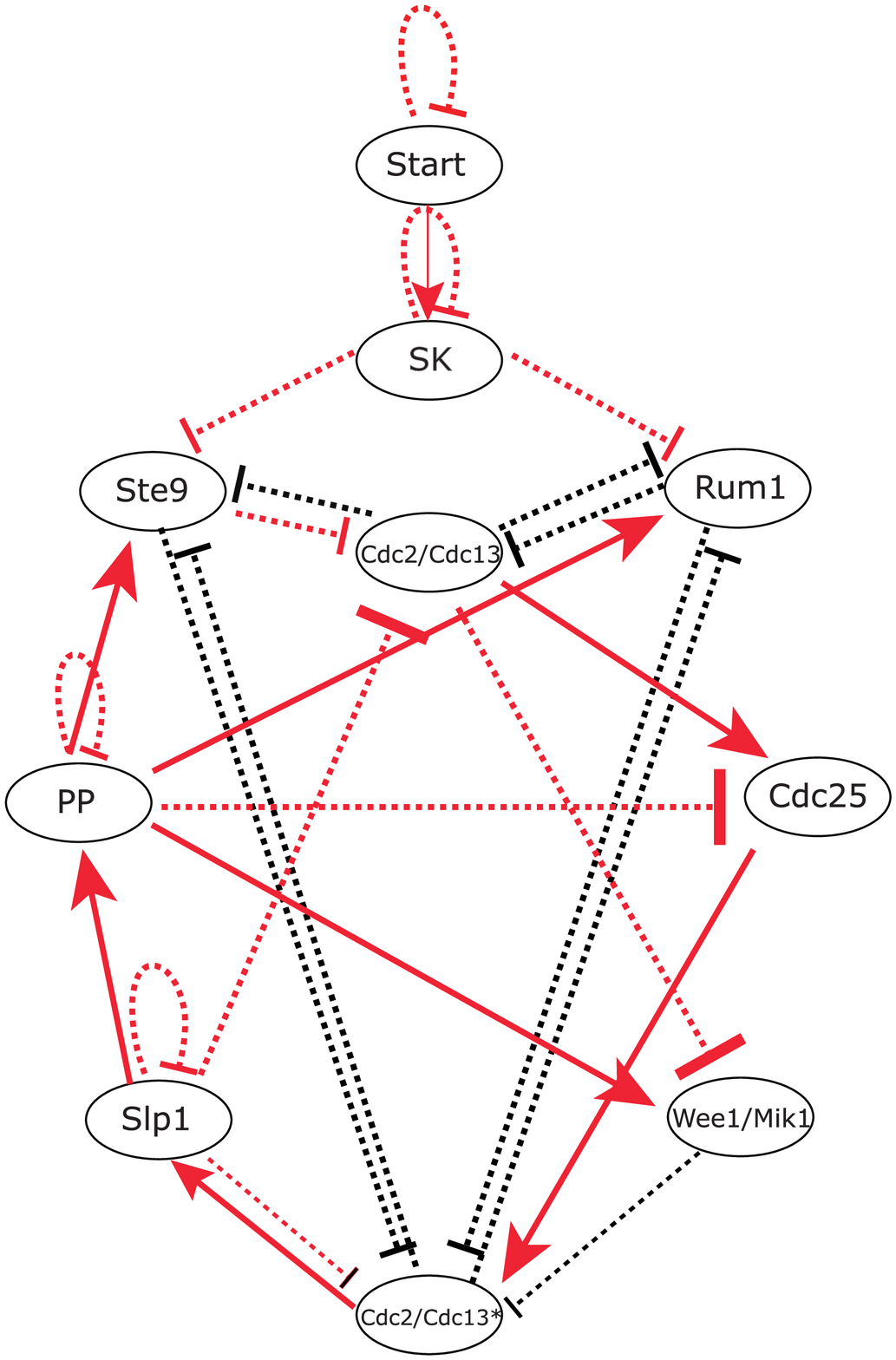}}
\caption{The minimal robustness backbone subnetworks of the fission yeast networks. All the links marked by red color contribute a backbone subnetwork, which is generated from the evaluation of robustness in Table.~\ref{table2} (also marked by red color). The remaining links form a residual auxiliary subnetwork. All the signs of the links are the same as in Fig.~\ref{fig:wiring}.   }\label{fig:backbonetwork}
\end{figure*}

We attempt to reconstruct a minimal but most robust backbone network of the fission yeast, in which we use the fewest edges but most of edges have highest scores of robustness, in the view of both potential and flux landscape. We show the backbone subnetwork in the Fig.~\ref{fig:backbonetwork} marked by red color. The 19-link backbone subnetwork is contributed following partly with Zeng's work\cite{Zeng2010}, in which they built a minimal subnetwork just to perform the fission yeast biological function.

We have highlighted the 19 edges in our robustness ranking table by red color (Table.~\ref{table2}). We can see that this subnetwork tends to choose those edges with high score of robustness.  This is to say, the minimal biological functional subnetwork tends to be a global stable one.

This result seems not so intuitive. As the backbone subnetwork directs to the minimal
network to perform the biological function, while the robust network means that it has
strong capability to persist in the mutation or fluctuation environments. Therefore, the minimal network should not always lead to the most robustness one.

In our study, we explore parameters listed in the Table.~\ref{table2} of both
potential and flux landscapes. The three elements which contribute the TES are
the key features representing the potential and flux landscape topography. Therefore, the rank of the TES is calculated in a quasi-quantitative way to gain insights on
both potential stability and period persistence for each edge. It suggests that due to  the consideration of flux landscape in the periodical dynamics, the minimal backbone
subnetwork with highest rank of TES tends to be a global robust one to perform the
cell cycle function. Therefore, we state that this study gives a physical principle and
basis in terms of the potential and flux landscape for the backbone finding.

 Furthermore, based on the global sensitivity analysis, we can identify key
links that change significantly the occupation probabilities of the states on the cell
cycle path and robustness ratio separating the dominant cycle loop from the rest
compared to the wild type. These links and nodes are responsible for biological
function of the cell cycle.

Furthermore, based on the global sensitivity analysis, we can identify key
links that change significantly the occupation probabilities of the states on the cell
cycle path and robustness ratio separating the dominant cycle loop from the rest
compared to the wild type. These links and nodes are responsible for biological
function of the cell cycle.

As we have stated, for this case of oscillation network, the flux landscape have a large impact on the dynamical behavior of the network. In Table.~\ref{table2}, if we delete the index of flux landscape, i.e.  $\bf RRflux$, there are several edges in the original network acutely changing their orders of importance. The edges such as  { $\bf SK$}  to $\bf Rum1$, and  $\bf Cdc25$ to { $\bf Cdc2/Cdc13*$} will lose their high ranks due to the missing of flux index, while the remaining edges such as { $\bf Cdc2/Cdc13$}  to $\bf Rum1$ and  to $\bf Ste9$  et. al. tend to squeeze into the subnetwork. This leads to the conclusion that the former edges tend to perform the biological cycle function while the remaining edges tend to keep the robustness of G1 state. It is in this perspective these results provide a strong support for the potential and flux landscape theory in the study of cell cycle.

Cell cycle is a hallmark of cancer. Cancer cells has a much faster speed of cell cycle than normal cells. Therefore, regulating the cell cycle speed is crucial for preventing and curing the cancer. From above global sensitivity analysis, we can identify the key nodes and links in the fission yeast cell cycle network for regulating the cell cycle speed. These key links and nodes form the backbone network of the cell
cycle. Therefore, we can based on this to do network design and network medicine discovery targeting the cancer.

\section*{Conclusions}

We explore the global natures of the networks. We found the network dynamics and global properties are determined by two essential features: the potential landscape and the flux landscape. While potential landscape quantifies the probabilities
of different states forming hills and valleys, the flux landscape quantifies the probability fluxes of different loops flowing through states. These two landscapes can be quantified through the decomposition of the dynamics into the detailed balance preserving part and detailed balance breaking part. While funneled landscape is crucial for the stability of the single attractor networks, the argument can be extended to the stabilities of the states on the oscillation paths by including them in the same (line) basin of attraction. Importantly, we have uncovered that the funneled flux landscape is crucial for the stable and robust oscillation flow.

This  provides a new interpretation of the origin of the limit cycle oscillations: There are always many cycles and loops forming the flux landscapes, each with a probability flux going through the loop. The oscillation only emerges when one specific loop stands out and carries much more probability flux than the rest of the others.

We studied the fission yeast cell cycle as an example to illustrate the idea. We found both the potential landscape and the flux landscape of the fission yeast cell cycle oscillations are funneled, which guarantees the global stability. While the funneled potential landscape guarantees the stabilities of the states on the oscillating path, the funneled flux landscape guarantees the directional flow of the oscillations which breaks the detailed balance and time reversal symmetry, leading to the stand out of the dominant flux loop against others. The stability and robustness of the oscillations are quantified through a dimensionless ratio of the steepness or gap versus the averaged variations or roughness of the landscape (measuring funnelness as we termed as robustness ratio RR).

We explore how RR changes with respect to the stimulations, self degradations, state switching rate or fluctuations, and changes in topology of the network (wirings). This allows us to identify the key factors and structure elements of the networks in determining the stability, speed and robustness of the fission yeast cell cycle oscillations.

Based on the global sensitivity analysis, we obtain that our most robust subnetwork is nearly the same as the minimal biological functional network, and by setting the cell cycle period as the evolution goal, we suggest the fission yeast should follow this evolution goal to form a 27-link network with faster period but not using minimal backbone network. We see that the non-equilibriumness characterized by the degree of detailed balance breaking from the energy pump quantified by the flux is the cause of  the energy dissipation for initiating and sustaining the replications essential for the origin and evolution of life. Finally we are looking forward to the good future by controlling the speed of the cell cycle as an important in designing targeting drugs for preventing and curing the cancer.

%

\end{document}